\documentclass[fleqn,usenatbib]{mnras}

\usepackage{newtxtext,newtxmath}
\usepackage{lastpage}  % 导言区加这个
\usepackage[T1]{fontenc}
\usepackage{graphicx}	% 插入图片
\usepackage{amsmath}	% 数学支持
\usepackage{booktabs}	% 更美观的表格线
\usepackage{rotating}	% 旋转表格
\usepackage{threeparttable} % 表格脚注
\usepackage{longtable}  % 长表格支持
\usepackage{color}      % 颜色支持
\usepackage{float}      % 图表位置
\usepackage{hyperref}
\usepackage{placeins}   % 控制浮动体位置
\usepackage{appendix}

\usepackage{CJKutf8}
\makeatletter
\def\endfigure{\end@float}
\makeatother
\DeclareRobustCommand{\VAN}[3]{#2}
\let\VANthebibliography\thebibliography
\def\thebibliography{\DeclareRobustCommand{\VAN}[3]{##3}\VANthebibliography}

\title[A New Algol-type Binary with an Accretion disc]{A New Algol-type Binary with an Accretion disc}

\author[T. He et al.]{
Tongyu He$^{1,2}$,
Jiao Li$^{2}$\footnotemark[2],
Xiaobin Zhang$^{3}$,
Mikhail Kovalev$^{2}$,
Zhibin Dai$^{2,4}$,
Zhenwei Li$^{2,4}$,\newauthor
Hongwei Ge$^{2,4,5}$,
Shunyi Lan$^{2}$,
Jiangdan Li$^{2,4}$,
Dengkai Jiang$^{2,4,5}$,\newauthor
Jianping Xiong$^{2,4}$,
Xuefei Chen$^{2,4,5}$,
Zhanwen Han$^{2,4,5}$\footnotemark[1]
\\
$^{1}$College of Physics Science and Technology, Hebei University, Baoding 071002, China\\
$^{2}$International Centre of Supernovae (ICESUN), Yunnan Key Laboratory of Supernova Research, Yunnan Observatories, CAS, Kunming 650216, China\\
$^{3}$Key Laboratory of Optical Astronomy, National Astronomical Observatories, Chinese Academy of Sciences, Beijing, 100012, China\\
$^{4}$Key Laboratory for the Structure and Evolution of Celestial Objects, CAS, Kunming 650216, China\\
$^{5}$University of Chinese Academy of Sciences, Beijing 100049, China
}

\begin{document}
\maketitle
% 在 maketitle 之后定义脚注文本，编号 1、2 对应上面 footnotemark[1]、[2]

%\label{firstpage}


\begin{abstract}
We present a comprehensive photometric and spectroscopic analysis of the Algol-type binary \textit{Gaia} DR3 1892576067672499328. We identified the system as a spectroscopic binary based on medium-resolution LAMOST spectra. Combined with \textit{TESS} photometry, we determine an orbital period of \( P = 2.47757 (1) \) days, a low mass ratio of \( q = 0.098 \pm 0.002 \), and an orbital inclination of \( i = 46.934^{+2.613}_{-1.11} \) degrees. The orbit is consistent with being circular (\( e = 0 \)). The binary comprises a \( M_1 = 1.817 ^{ +0.106}_{-0.202} \,M_\odot \), \( R_1 = 1.265^{+0.121}_{-0.160}\,R_\odot \) A-type primary and a Roche-lobe-filling secondary of \( M_2 = 0.179 ^{ +0.011}_{-0.020} \,M_\odot \), \( R_2 = 1.994 ^{ +0.041}_{-0.077} \,R_\odot \). The double-peak H$\alpha$ emission line indicates the possible existence of a Keplerian accretion disc. We established a simple standard accretion disc model and modeled the geometric and dynamical properties of the accretion disc. The obtained outer disc radius $R_{\mathrm{out}} \approx 3.36 \pm 0.43\,R_\odot$ is consistent with the values inferred from the emission velocity of H$\alpha$. Systemic velocity variations observed over time suggest the possible presence of a tertiary companion, with a minimum mass of $M_3 > 0.369 \pm 0.024 \,M_\odot$. Given the low mass ratio, the secondary may evolve into a proto-helium white dwarf, forming an \text{EL CVn}-type system in the future. This system offers valuable insights into accretion dynamics and the formation of binaries.
\end{abstract}


%\keywords{Triple stars (1714);  Algol variable stars(24); Stellar accretion(1578); Radial velocity (1332); Light curves(918); Spectroscopy (1558)}           
 %%%%%%%%%%%%  介绍
\section{Introduction} 
\label{sec:intro} 

Nearly half of the stars in the Milky Way are members of binary or higher-order multiple systems, rather than evolving in isolation. Such systems play a fundamental role in understanding stellar formation, binary interaction processes, and the overall dynamical evolution of galaxies \citep{2008eggleton, 2010eggleton, He2023}. Among them, Algol-type semi-detached binaries are particularly valuable for investigating mass transfer, accretion phenomena, and long-term evolutionary outcomes in close binary systems \citep{Paczynski1971, Hall1989}.

Typically, an Algol-type binary consists of a main-sequence primary—most commonly of spectral type B, A, or F—and an evolved subgiant or giant secondary that fills its Roche lobe. These systems frequently exhibit the so-called “evolutionary reversal”, in which the less massive secondary appears more evolved than the primary due to prior episodes of mass transfer. Many Algol-type binaries remain in active mass transfer stages, exhibiting diverse photometric and spectroscopic behavior, including orbital period variations and characteristic double-peaked H$\alpha$ emission lines indicative of accretion discs \citep{1996ApJR, Richards1999, Olson2011}.

The prevailing theoretical framework for interpreting such accretion features originated from the pioneering work of Huang and Struve \citep{1941ApJ, 1957AJ, 1956ApJH}. They proposed that gas flowing through the inner Lagrangian point (L1) can either directly impact the surface of the primary star, or, in systems with longer orbital periods, form a stable accretion disc due to the conservation of angular momentum. In such wide-separation systems, the stream trajectory does not directly impact the surface of the primary, as the stellar radius is smaller than the minimum distance of the stream. This condition favors the formation of an accretion disc, making disc-like structures more likely to develop and more easily detectable in observations. Observational surveys support this theoretical picture. \cite{1976IAUS} detected H$\alpha$ emission in 25 of 40 surveyed Algol systems, and \cite{1980IAUS} later extended the sample to over 50, concluding that the emission strength and stability are positively correlated with the orbital period. Short-period systems like \mbox{RW~Tau} and \mbox{U~Cep}, by contrast, show only transient or weak features. By analyzing the H$\alpha$ differential spectra, \cite{Richards1999} discovered that there are four basic morphological types of accretion structures in the Algol-type binary stars:% (Figure \ref{figure 9}):%These results suggest that the morphology of the accretion structure is related to the orbital period \citep{Richards1999}:

%\begin{figure}[ht!]
%\centering
%\includegraphics[width=11cm]{9.png}
%\caption{Schematic illustrations of four types of H$\alpha$ emission profiles and their corresponding accretion structures in Algol-type binary systems.
%\label{figure 9}}
%\end{figure}

1. Double-peaked emission systems, characterized by an accretion structure in the form of either a classical or transient accretion disc;

2. Single-peaked emission systems, in which the infalling material follows a curved stream trajectory and partially encircles the primary, forming an accretion ring-like structure between the binary components;

3. Alternating single- and double-peaked emission systems, which exhibit a transition from single-peaked to double-peaked profiles within a single orbital cycle. This behavior suggests a dynamic accretion structure, likely caused by phase-dependent changes in the visibility or geometry of the gas stream and disc. The variation may result from fluctuating mass transfer rates, transient disc formation, or projection effects due to orbital motion.

4. Weak-emission systems, in which there is little to no evidence for the presence of any accretion structure at all orbital phases, as indicated by the consistently weak residual spectra \citep{1996ApJR}.

%1. Long-period systems ($P \geq 6$ days): Stable, classical accretion 
%discs form, where gas orbits the primary star and gradually spirals inward due to viscous effects. These discs produce strong, symmetric, double-peaked H$\alpha$ emission lines, resembling those observed in cataclysmic variables or low-mass X-ray binaries.
    
%2. Intermediate-period systems ($4.5 < P < 6$ days): Transient and unstable discs may form intermittently, producing weaker, asymmetric emission features \citep{1982ApJK, 1989ApK}.
    
%3. Short-period systems ($P < 4.5$ days): The accretion stream is more likely to impact the stellar surface directly, leading to the formation of hotspots or incomplete rings with little or no pronounced H$\alpha$ emission \citep{1996ApJR}.

The morphology of accretion structures is primarily determined by the system’s mass ratio, orbital period, and the size of the accretor. According to the theoretical framework of \cite{1975AL}, if the primary star's radius is smaller than a critical threshold, the stream possesses sufficient angular momentum to avoid direct impact and form a disc. Conversely, if the star is large enough to intercept the stream, a direct collision occurs, leading to the formation of a hot impact region and possibly a transient ring-like structure around the star. The morphology, stability, and evolution of these accretion structures are further governed by the stream's hydrodynamics and thermodynamics, including its kinetic energy, entropy, Coriolis deflection, and the rotational properties of the accreting star.

More complex dynamics arise in hierarchical triple systems, where an inner binary is orbited by a distant tertiary component. These systems consist of a close inner binary orbiting a distant tertiary, typically arranged in a dynamically stable configuration \citep{aa}. In such configurations, the outer tertiary orbits the barycenter of the inner binary, allowing the system to be approximated as a nested set of Keplerian orbits \citep{He2023}. The presence of a third body may induce orbital period modulations via the light-travel time effect or dynamical interactions, and may even trigger the Kozai–Lidov mechanism, enhancing mass transfer instability \citep{Tokovinin1997, 2016Borkovits, Naoz2016}. These multi-body effects can significantly impact the long-term evolution and morphology of accretion structures.

From a stellar evolution perspective, the study of Algol-type binaries carries far-reaching implications. Mass-losing secondaries may eventually evolve into extremely low-mass helium white dwarfs (ELM WDs), which can lead to EL~CVn-type systems, many of which have been identified in large-scale surveys, such as WASP and \textit{Kepler} \citep{Maxted2014, Istrate2014, Li2019}. These systems are also considered potential gravitational wave sources for future space-based detectors such as LISA \citep{Burdge2023}. Algol systems can be used to test stellar evolution models \citep{2020hanz}, study post-mass-transfer products, and conduct asteroseismic investigations \citep{2024A&AV, 2013ApJZ}. As such, Algol-type binaries play an important role in our understanding of stellar physics, binary evolution, and accretion phenomena.

%%%%%%%%%%%%%%%%%%%%%%%%%%%%%%%%%%%%%%%%%%%%%%

In this study, we analyze a short-period Algol-type binary system ($Gaia$ DR3 1892576067672499328) with an accretion disc. Section~\ref{sec:2} presents the observational data, including both spectroscopic and photometric data. In Section~\ref{sec:3}, we describe our analysis methods, with a particular focus on radial velocity (RV) curve fitting and light curve modeling. Section~\ref{sec:4.1} discusses the H$\alpha$ emission features, presenting evidence for the presence of an accretion disc and simulations of its possible structure. Section~\ref{sec:4.2} explores the potential existence of a tertiary companion in the binary system. Section~\ref{sec:4.3} examines the future evolutionary pathway of the binary. A summary of our findings is provided in Section~\ref{sec:5}.

\section{Observation Data} \label{sec:2}
\subsection{Spectroscopic Observations}

The Large Sky Area Multi-Object Fibre Spectroscopic Telescope (LAMOST), situated in Hebei, China, is a quasi-meridian 4-meter reflecting Schmidt telescope designed for wide-field spectroscopic surveys. It offers a wide field of view of $5^\circ$ and is equipped with 4,000 optical fibers, allowing for the simultaneous acquisition of spectra for up to 4,000 objects over a 20 square degree area. LAMOST has conducted large-scale spectroscopic observations across the northern sky, providing valuable data that have greatly advanced our understanding of the Galaxy. The twelfth data release of the LAMOST Medium-Resolution Spectroscopy survey (LAMOST-MRS DR12) provides a big dataset containing detailed spectroscopic information for tens of millions of stars, including radial velocities and various stellar parameters. The LAMOST-MRS instrument consists of two spectral arms: the red arm covers the wavelength range from 6300 to 6800\,\AA, while the blue arm spans 4950 to 5350\,\AA.

%%%%%%%%%%%%%%%%%%%%%%%%%%%%%%%%%%%%%%%%%%%%%%%%%%%%%%%%%%%%%%%%%%
The spectroscopic observations of this system were obtained from the LAMOST-MRS DR12. The target was observed on 14 separate nights between August 16, 2018, and September 4, 2021. Given the relatively short orbital period ($P = 2.47757$\,days), the 20-minute exposure time of LAMOST observations offers sufficient phase coverage to resolve orbital motion. We selected spectra with a signal-to-noise ratio (S/N) $\geq 20$ in the blue arm for radial velocitys (RVs) analysis \citep{li-2021}. A total of 45 blue-arm and 60 red-arm spectra were used.

Since the blue arm contains a richer set of metal absorption lines, it is usually used for Radial velocity 
(RV) extraction. We derived the radial velocities of the two components using the classical cross-correlation 
function (CCF) method, cross-correlating the observed spectra $f(x)$ with a synthetic template $g(x)$. The 
template was generated using the \texttt{Spectrum} software with stellar parameters of $T_{\rm eff} = 5775$\,K and $\log g = 4.0$  \citep{spectra,2014ispec,2019Blanco}. It is defined as:

\begin{equation}
\mathrm{CCF}(h) = \int_{-\infty}^{+\infty} f(x)\, g(x + h)\, dx,
\end{equation}

where $h$ represents the Doppler shift in units of km/s. The CCF is normalized between $-1$ and $+1$, with 
values near $+1$ indicating strong correlation. This method, widely used for spectroscopic binary analysis 
\citep{2017mer,li-2021, He2023,He2024}, also enables the identification of multiple-lined features. The 
cross-correlation analysis of the observed spectra was carried out using the \texttt{laspec} software package 
\citep{2020zhangbo, 2021zhangbo}. Table~\ref{tab:rv} presents the resulting radial velocities (RVs) along with their corresponding modified Julian dates (MJDs).

%A detailed analysis of the H$\alpha$ emission profile in the red-arm spectra, along with its implications for the presence of disc structures, is presented in Section~\ref{sec:3.3}.

\subsection{Photometric Observations}

In recent years, NASA’s Transiting Exoplanet Survey Satellite (\textit{TESS}) has become an important resource for identifying binary star systems. Although its primary mission focuses on exoplanet detection, 
\textit{TESS} also captures high-precision photometric data that reveal brightness variations caused by 
stellar multiplicity. The resulting light curves are instrumental in analyzing binaries that exhibit periodic 
flux changes due to eclipses, tidal distortion, or reflection effects, thus offering a strong basis for 
further analysis \citep{Ricker2015, Sullivan2015}. Complementing this, the Zwicky Transient Facility (ZTF) is 
a high-cadence, ground-based optical survey covering the northern sky in the $g$, $r$, and $i$ bands. Its 
broad sky coverage and frequent observations make it particularly effective for tracking long-term photometric variability in binary systems \citep{Bellm2019, Masci2019}.

We collected and analyzed the light curve data of the binary system \textit{Gaia} DR3 1892576067672499328. The photometric data were obtained from the \textit{TESS} through the Quick Look Pipeline (QLP). The target was 
observed in two separate \textit{TESS} sectors—Sector 56 and Sector 83—yielding two segments of light curves 
with good temporal coverage. The \textit{TESS} light curves were processed using the \texttt{lightkurve} 
Python package \citep{2018ascl.soft12013L}, following standard calibration procedures, including outlier 
removal and normalization. The combined dataset spans over 50 days, suitable for precisely constraining 
orbital phase variations. Additionally, we identified photometric data for the same target from the ZTF, which cover observations in the $g$ and $r$ bands. The ZTF dataset extends the 
photometric baseline to over five years, effectively complementing the shorter-duration \textit{TESS} observations. To ensure data quality, outliers were removed from the ZTF light curves prior to analysis.

To determine the binary system's orbital period, we performed a Lomb–Scargle periodogram analysis \citep{1982Scargle, 2018ascl.soft12013L} using the cleaned \textit{TESS} photometry only.\footnote{We find a 
contaminating star is near this system, and the large pixel scale of \textit{TESS} likely includes flux 
contributions from this nearby object. We corrected the \textit{TESS} flux using the formula from \cite{2012PASP..124..985S}.} The resulting best-fit period is $P = 2.47757(1)\ \mathrm{days}$. This TESS-based period is adopted throughout the paper for all subsequent photometric and spectroscopic analyses. 

%To determine the binary system's orbital period, we performed a Lomb–Scargle periodogram analysis \citep{1982Scargle, 2018ascl.soft12013L} on the cleaned \textit{TESS} photometry. The resulting best-fit period is $P = 2.47757 \pm 0.00001\ \mathrm{days}$, which yields a well-defined phase-folded light curve with no evidence of significant aliasing or secondary periodicities. This period is adopted throughout the paper for all subsequent photometric and spectroscopic analyses.

%%%%%%%%%%%%%%%%%

\section{Methods and analysis} 
\label{sec:3}

In the previous section, we presented the spectroscopic and photometric observations of the binary $Gaia$ DR3 
1892576067672499328. In this section, we perform a comprehensive analysis, including Section~\ref{sec:3.1}, 
where we model the radial velocity curves, and Section~\ref{sec:3.2}, which presents the light curve fitting.%; Section~\ref{sec:3.3}, focusing on the H$\alpha$ emission profiles and the accretion disc; and Section~\ref{sec:3.4}, which investigates the possible presence of a third body in the system.

\begin{figure}
\centering
\includegraphics[width=0.5\textwidth]{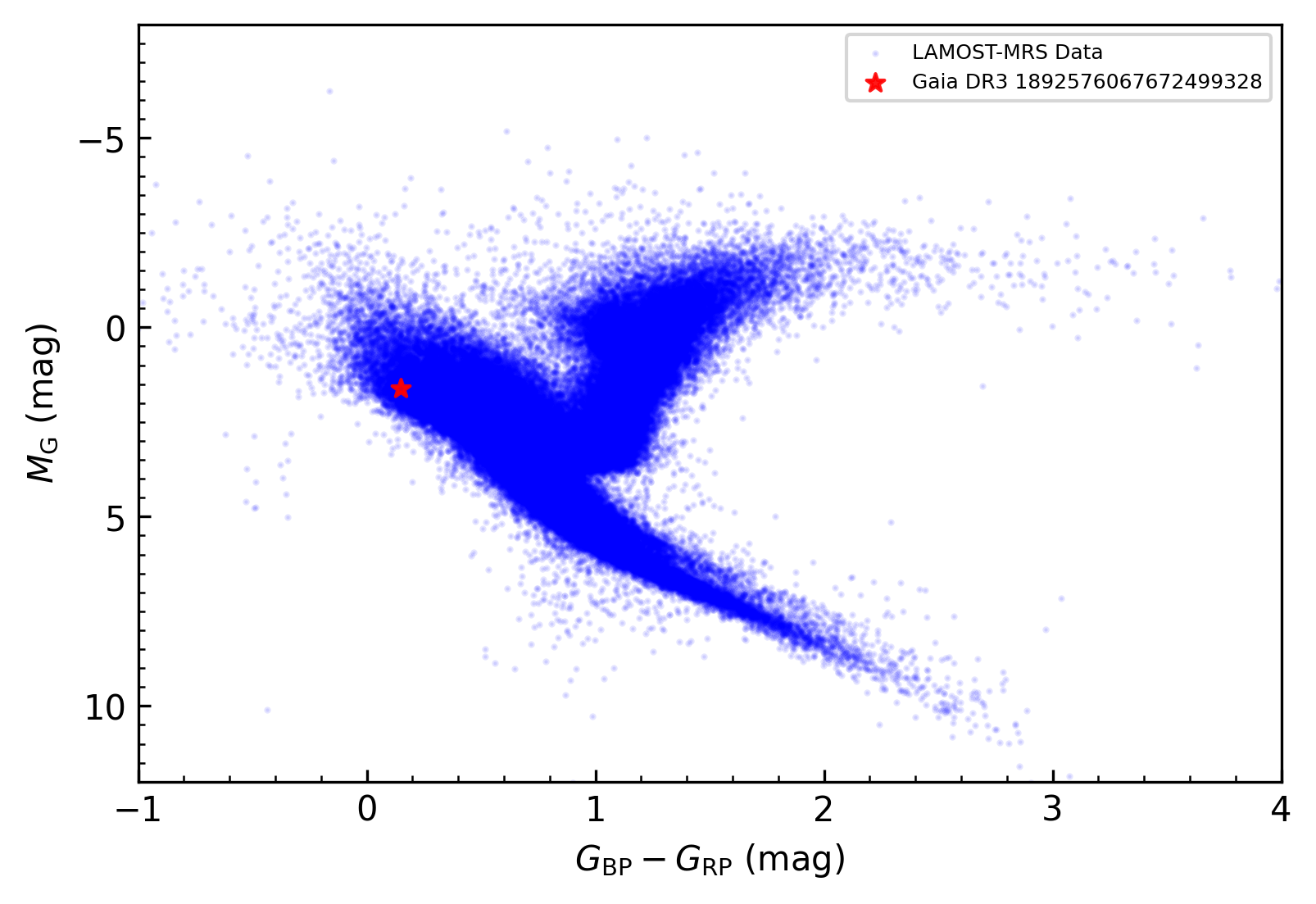}

\caption{The color--magnitude diagram shows our target star (red star) in comparison to the LAMOST sample 
(blue dots). The horizontal axis represents the color index $G_\mathrm{BP} - G_\mathrm{RP}$, while the 
vertical axis indicates the absolute magnitude $M_\mathrm{G}$. All photometric measurements are taken from \textit{Gaia} DR3.
}
\label{figure1}
\end{figure}

Figure~\ref{figure1} presents the color--magnitude diagram (CMD) for Gaia DR3 1892576067672499328. Blue dots 
represent the full sample of stars observed with LAMOST-MRS, while the red star marks the position of our target. The plot reveals that the sample we filtered is situated close to the main sequence.

\subsection{Radial Velocity Analysis}
\label{sec:3.1}
%Two distinct absorption components are clearly visible in most spectra, allowing reliable determination of the line-of-sight velocities for both the primary and the secondary components. To investigate the orbital motion and line profile variability in detail, we constructed a dynamical spectrum around the Mgb triplet region, centered at a rest-frame wavelength of $H_0 = 5185.04$\,\AA. This region was chosen because it contains strong and sharp absorption features from both components of the binary, which are clearly separable across most orbital phases. We first plot all available spectra in velocity space to illustrate the Mgb line profiles at different epochs (see Figure~\ref{figure 2-}). Then, to visualize their evolution with orbital phase, we binned the spectra into 50 phase intervals and averaged them, resulting in the two-dimensional dynamical spectrum (Figure~\ref{figure 2-}). The sinusoidal motion of both components is clearly seen, confirming their SB2 nature and consistent with the orbital solution.
Using the radial velocity (RV) measurements listed in Table~\ref{tab:rv}, we find that the system is a typical double-lined spectroscopic binary (SB2). In most spectra, two distinct absorption components are clearly 
visible, allowing for reliable determination of the velocities of both the primary and secondary stars. For 
example, in the blue-arm spectra near the Mg\text{b} triplet, we observe two deep absorption lines in the 
two-dimensional dynamical spectra that vary systematically with orbital phase, reflecting the orbital motion 
of both stars. The RVs of the two stars are clearly detected, confirming it's a SB2 binary and showing excellent agreement with the orbital solution (see Figure~\ref{figure2-} for more details).

\begin{figure}
\centering
\includegraphics[width=0.5\textwidth]{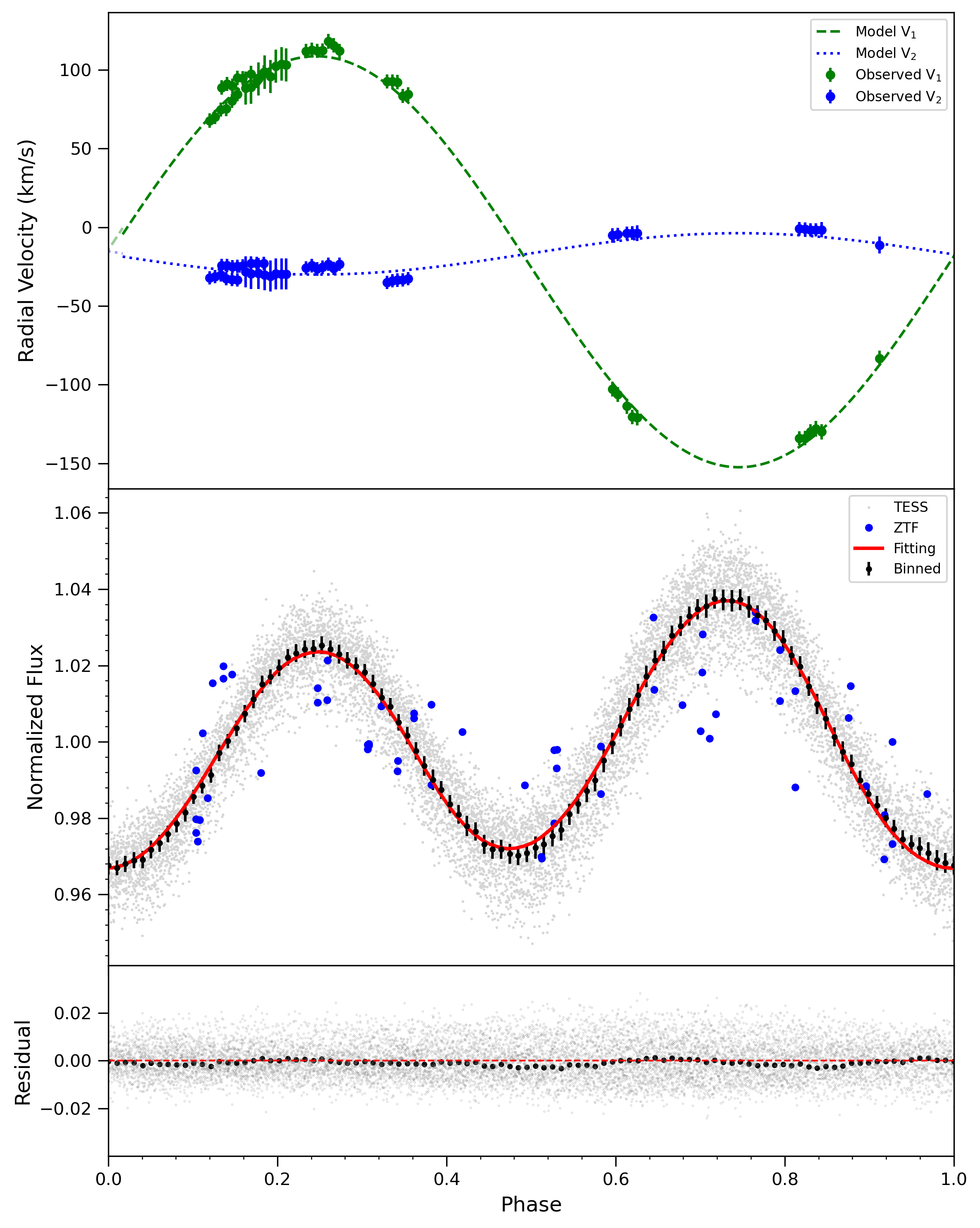}

\caption{Phase-folded observational data and best-fit model for binary $Gaia$ DR3 1892576067672499328. Top 
panel: Radial velocity curves. Green and blue symbols represent the observed RVs of the primary and secondary 
stars, respectively. The dashed and dotted lines show the corresponding model predictions. middle panel: 
Phase-folded light curves. Black and blue points represent flux measurements from \textit{TESS} and ZTF, 
respectively, while the red line indicates the best-fitting model. Bottom panel: Residuals between the observed light curves and the fitted model.
\label{figure3}
}
\end{figure}

We phase-folded the RVs using the orbital period derived from the \textit{TESS} light curve, and constructed a phase-RV diagram (Figure~\ref{figure3}). The orbital parameters were obtained by fitting the RVs with a 
Keplerian SB2 model using the \texttt{RadVel} package \citep{2018Fulton,2024lijiao}. The model is described by:

\begin{align}
RV_1(t) &= \gamma + K_1 \left[\cos(\theta(t) + \omega) + e \cos \omega\right], \\
RV_2(t) &= \gamma - K_2 \left[\cos(\theta(t) + \omega) + e \cos \omega\right],
\end{align}

where $\gamma$ is the systemic velocity, $K_1$ and $K_2$ are the semi-amplitudes of the primary and secondary 
star, respectively, $e$ is the orbital eccentricity, $\omega$ is the argument of periastron, and $\theta(t)$ is the true anomaly computed from Kepler's equation.

The best-fit parameters, including $K_1$, $K_2$, $\gamma$, $e$, and the mass ratio $q = K_1 / K_2$, were 
derived using an MCMC fitting procedure \citep{2013Foreman}. The resulting RV curves are shown at the top of 
Figure~\ref{figure3}; green and blue symbols represent the observed RVs of the primary and secondary stars, 
respectively. The dashed and dotted lines show the corresponding model predictions. We can determine some 
orbital parameters of the binary. The parameters are as follows: \(q = 0.098 \pm 0.002\), \(K_2 = 
130.366^{+0.06}_{-0.061}\), \(\sqrt{e}\cos\omega_1 = 0\), \(\sqrt{e}\sin\omega_1 = 0 \). A spectral synthesis 
analysis using the method of \cite{Mikhail2022} indicates an effective temperature of approximately $7768 \pm 
307 $ K for the primary star. Projected rotational velocities ($ v\sin i$) for both components were also 
estimated from the observed line broadening. $v \sin i$ of the primary star is $20 \pm 9 $ \text{km/s} and the $ v \sin i$ of the second star is $32 \pm 10 $ \text{km/s}. %These spectroscopic constraints serve as essential priors for the light curve and structural modeling described in later sections.

\subsection{Light curve Analysis}
\label{sec:3.2}

The orbital period of the binary system was determined to be 2.47757 days based on \textit{TESS} observations \citep{2017Lomb}. We phase-folded the light curve using this period, as shown in the middle panel of 
Figure~\ref{figure3}, where the blue and black points represent the observed photometric data of the two components. 

%We find a contaminating star is located within $10''$ of this system, and the large pixel scale of \textit{TESS} likely includes flux contributions from this nearby object. To correct for this contamination, we first analyzed ZTF data to estimate the contaminating star’s brightness and variability. We found that the contaminant does not show any significant periodic modulation. Next, we corrected the \textit{TESS} flux using the formula from \cite{2012PASP..124..985S}.

%\begin{equation}
%\mathrm{flux}_{\mathrm{corrected}} = \frac{\mathrm{flux}_{\mathrm{norm}} - 1 + f}{f},
%\end{equation}
%where $f$ is the flux fraction of the star.

We then modeled the \textit{TESS} light curve using the PHOEBE code (Physics of Eclipsing Binaries; \cite{2005Pr, 2016Pr,2020Jones}), which is based on the Wilson-Devinney (WD) framework \citep{1971Wilson}. The light curve shows an ellipsoidal modulation with two unequal maxima: the brightness at phase $\phi \approx 
0.25$ is significantly lower than at $\phi \approx 0.75$, with the difference reaching several percent.

Initially, we considered whether relativistic Doppler boosting could account for this asymmetry. However, the 
expected effect is extremely small and cannot account for the observed difference. We therefore introduced a 
\textit{hotspot} on the surface of the primary star in PHOEBE to account for the asymmetry. The addition of 
this feature led to a significantly improved fit. The hotspot may correspond to an accretion impact region on 
the stellar surface due to material from an accretion structure.

We then used a Markov Chain Monte Carlo (MCMC) sampler within PHOEBE to explore the parameter space and 
determine the best-fit values with uncertainties \citep{2013Foreman}. Table~\ref{tab:model_priors} summarizes the priors adopted for each parameter in the binary model.

\begin{table*}
\centering
\renewcommand{\arraystretch}{1.3}  % 调整行距为默认的1.3倍
\caption{Model parameters and their prior \textbf{distributions. }}
\label{tab:model_priors}
\begin{tabular}{lll}
\hline
\textbf{Parameter} & \textbf{Prior} & \textbf{Description / Units} \\
\hline
Semi-major axis $a$             & Uniform [5, 30]                & $R_\odot$ \\
Primary radius $R_1$            & Uniform [0.02, 3.0]            & $R_\odot$ \\
Temperature ratio $T_2/T_1$     & Uniform [0, 1]                 & -- \\
Third light fraction $L_3$      & Uniform [0, 0.5]               & Fraction of total system flux \\
Inclination $i$                 & Uniform [$10^\circ$, $90^\circ$]   & Degrees \\
Hotspot temperature ratio       & Uniform [1.0, 1.55]             & $T_{\mathrm{spot}}/T_1$ \\
Hotspot radius                  & Uniform [$5^\circ$, $20^\circ$]     & Degrees (Open angle on the stellar surface) \\
Hotspot longitude               & Uniform [$0^\circ$, $360^\circ$]     & Degrees \\
Mass ratio $q$                  & $\mathcal{N}(0.098, 0.002)$   & -- \\
Primary mass $M_1$              & Uniform [0.02, 3]              & $M_\odot$ \\
\hline
\end{tabular}
\end{table*}

We fixed the effective temperature of the primary star to the spectroscopic value derived in Section~\ref{sec:3.1} ($T_1 = 7768 \pm 307 $ K), and incorporated both the spectroscopic mass ratio q and the 
RV semi-amplitude of the secondary \(K_2 = 130.366^{+0.06}_{-0.061}\) \text{km/s} as priors in the MCMC fit. 
The sampling was performed using 40 walkers over 5000 steps. We adopted the median of the posterior distribution as the best-fit value, and the standard deviation as the 1$\sigma$ uncertainty.

The final derived parameters include the orbital inclination, temperature ratio, scaled radii $R_1/a$ and $R_2/a$, the third light contribution, the semi-major axis and primary mass $M_1$. During the light-curve modelling,  we included the third light contribution ($L_3$). This was motivated by the fact that the observed light curve cannot be satisfactorily reproduced without including an additional light contribution, and the fit quality improves significantly when this parameter is allowed to vary. A possible explanation for this contribution, such as a tertiary component, is discussed in Section \ref{sec:4.2}. The model fit is shown in the middle panel of Figure~\ref{figure3}, where the black points indicate the folded \textit{TESS} data and the red line shows the best-fit model. Residuals are plotted in the lower panel. The best-fit parameters for 
the star (\textit{Gaia} DR3 1892576067672499328) are as follows: 
$a = 9.697^{+0.172}_{-0.385}\ R_\odot$, 
$R_1 = 1.265^{+0.121}_{-0.160}\ R_\odot$, 
$T_2/T_1 = 0.712^{+0.048}_{-0.031}$, 
$L_3 = 0.205^{+0.057}_{-0.051}$, 
$i = 46.934^{+2.613}_{-1.110}\ ^\circ$, 
$q = 0.098 \pm 0.002$, 
$M_1 = 1.817^{+0.106}_{-0.202}\ M_\odot$, 
hotspot temperature ratio $T_\mathrm{spot}/T_1 = 1.43^{+0.050}_{-0.040}$, 
hotspot radius $= 11.393^{+1.035}_{-0.616}\ ^\circ$, 
and hotspot longitude $= 127.504^{+1.480}_{-0.529}\ ^\circ$.

Light curve modeling indicates that the system is a low mass-ratio Algol-type binary, in which the primary is 
an A-type star and the secondary is a subgiant that has completely filled its Roche lobe. The temperatures of 
the primary star and the secondary star are 7768$ \pm 307 $K and $5531 ^{+432}_{-325} $K respectively, and the mass ratio is 0.098. Based on the results from Sections \ref{sec:3.1} and \ref{sec:3.2}, where we analyzed the radial velocity curves and light curves respectively, we have derived key orbital and physical parameters of 
the binary system. With these parameters in hand, we can further determine fundamental stellar quantities by 
applying Kepler's third law and the following standard physical relations.

To compute the surface gravity ($\log g$) of the binary components, we use:
\begin{align}
\log g &= \log \left( \frac{G M}{R^2} \right) ,
\end{align}
where $G$ is the gravitational constant in cgs units (cm$^3$ g$^{-1}$ s$^{-2}$), and $M$ and $R$ are the stellar mass and radius, respectively. The luminosity is given by:
\begin{align}
\frac{L}{L_\odot} \approx \left( \frac{R}{R_\odot} \right)^2 \left( \frac{T_{\text{eff}}}{T_\odot} \right)^4 ,
\end{align}
where $T_{\text{eff}}$ is the effective temperature, and $L_\odot$, $R_\odot$, and $T_\odot$ are the solar luminosity, radius, and temperature, respectively. The Roche lobe radius for the secondary star can be approximated by the empirical relation:
\begin{equation}
\frac{R_{\text{L}}}{a} \equiv x_{\text{L}}(q) \approx \frac{0.49 q^{2/3} + 0.27q - 0.12q^{4/3}}{0.6q^{2/3} + \ln(1 + q^{1/3})}, \quad q \leq 1,
\end{equation}
where $a$ is the orbital semi-major axis, and $q = M_2/M_1$ is the mass ratio of the system \citep{1983ApJ}. For detailed parameters, please refer to Table \ref{tab:physical_params} about the parameters of the semi-detached binary star \textit{Gaia} DR3 1892576067672499328. The parameters were determined through combined modeling of the 
\textit{TESS} light curve and radial velocity data. The table includes fundamental stellar parameters (masses, radii, temperatures, surface gravities, and luminosities), orbital elements (inclination, semi-major axis, 
mass ratio), properties of the hotspot used to model light curve asymmetry, and the fractional contribution of the third light. %The minimum mass of a tertiary component is also estimated based on the O$-$C analysis, assuming an orbital inclination of $90^\circ$.

%%%%%%%%%%%%%%%%%%%%%%%%%%%%%%%%%%%%%%%%
\begin{table}
\centering
\renewcommand{\arraystretch}{1.3}  % 调整行距为默认的1.3倍
\caption{Physical and orbital parameters of the star \textit{Gaia} DR3 1892576067672499328.}
\label{tab:physical_params}
\begin{tabular}{lcl}
\hline
Parameter & Symbol & Value  \\
\hline

Orbital inclination & $i$ & $46.934^{+2.613}_{-1.110} \ ^\circ$  \\
Inner Period & P$_\text{{in}}$ & $2.47757 (1)$ days \\
Third light fraction & $L_3$ & $0.205^{+0.057}_{-0.051}$  \\
Temperature ratio (Hotspot) & $T_2/T_1$ & $1.43^{+0.050}_{-0.040}$  \\
Hotspot radius & $r_{\mathrm{spot}}$ & $11.393^{+1.035}_{-0.616}\ ^\circ$  \\
Hotspot longitude & $\phi_{\mathrm{spot}}$ & $127.504^{+1.480}_{-0.529} \ ^\circ$  \\
Mass ratio & $q$ & $0.098 \pm 0.002$  \\
Radius (primary) & $R_2$ & $1.994^{+0.041}_{-0.077}\ R_\odot$ \\
Radius (secondary) & $R_1$ & $1.265^{+0.121}_{-0.160}\ R_\odot$ \\
Mass (primary) & $M_1$ & $1.817^{+0.106}_{-0.202}\ M_\odot$  \\
Mass (secondary) & $M_2$ & $0.179 ^{+0.011}_{-0.020}\ M_\odot$  \\
Semi-major axis & $a$ & $9.697^{+0.172}_{-0.385}\ R_\odot$  \\
Orbital radius (primary) & $a_1$ & $0.866 ^{+0.023}_{-0.033}\ R_\odot$  \\
Orbital radius (secondary) & $a_2$ & $8.822 ^{+0.162}_{-0.345}\ R_\odot$  \\
Roche lobe radius (primary) & $R_{\mathrm{L1}}$ & $5.614 ^{+0.104}_{-0.222}\ R_\odot$  \\
Roche lobe radius (secondary) & $R_{\mathrm{L2}}$ & $1.994 ^{+0.041}_{-0.077}\ R_\odot$ \\
Effective temperature (primary) & $T_1$ & $7768 \pm 307\ \mathrm{K}$  \\
Effective temperature (secondary) & $T_2$ & $5531 ^{+432}_{-325}\ \mathrm{K}$ \\
Projected rotation velocity (primary) & $v \sin i_1$ & $20 \pm 9\ \mathrm{km\ s^{-1}}$  \\
Projected rotation velocity (secondary) & $v \sin i_2$ & $32 \pm 10\ \mathrm{km\ s^{-1}}$  \\
Primary luminosity & $L_1$ & $5.113 \pm 1.476\ L_\odot$  \\
Secondary luminosity & $L_2$ & $3.363 \pm 1.260\ L_\odot$ \\
Surface gravity (primary) & $\log g_1$ & $4.501 \pm 0.122$ [cgs]  \\
Surface gravity (secondary) & $\log g_2$ & $3.088 \pm 0.018$ [cgs]  \\

Minimum mass of tertiary component & $M_3$ (min) & $0.369 \pm 0.024\ M_\odot$  \\
\hline
\end{tabular}
\end{table}

\section{Result and Discussion} 
\label{sec:4}

In Section~\ref{sec:3.1}, we model the radial velocity curves and derive key orbital parameters of the system, such as the mass ratio, velocity semi-amplitude, and orbital eccentricity. Section~\ref{sec:3.2} presents the light curve fitting, where we analyze the \textit{TESS} 
light curve to obtain the physical parameters of the binary. Our modeling results indicate that the binary system \textit{Gaia} DR3 1892576067672499328 is an Algol-type semi-detached binary, in which the less massive secondary star fills its Roche lobe and transfers mass to the more 
massive primary star. The primary is an A-type main-sequence star, while the secondary is a subgiant. For detailed parameters, please refer to the Table \ref{tab:physical_params}. 

We also investigated additional properties of the binary system \textit{Gaia} DR3 1892576067672499328, 
including the presence of an accretion disc and a potential tertiary companion. Section~\ref{sec:4.1} presents the H$\alpha$ emission features and provides both observational evidence and simulated models for the 
accretion disc structure. In Section~\ref{sec:4.2}, we explore the possible existence of a third companion. 
Finally, Section~\ref{sec:4.3} discusses the future evolutionary pathways of the system.

\subsection{\texorpdfstring{H$\alpha$ Profiles and Accretion disc}{Hα Profiles and Accretion disc}}
\label{sec:4.1}

\begin{figure}
\centering
\includegraphics[width=0.5\textwidth]{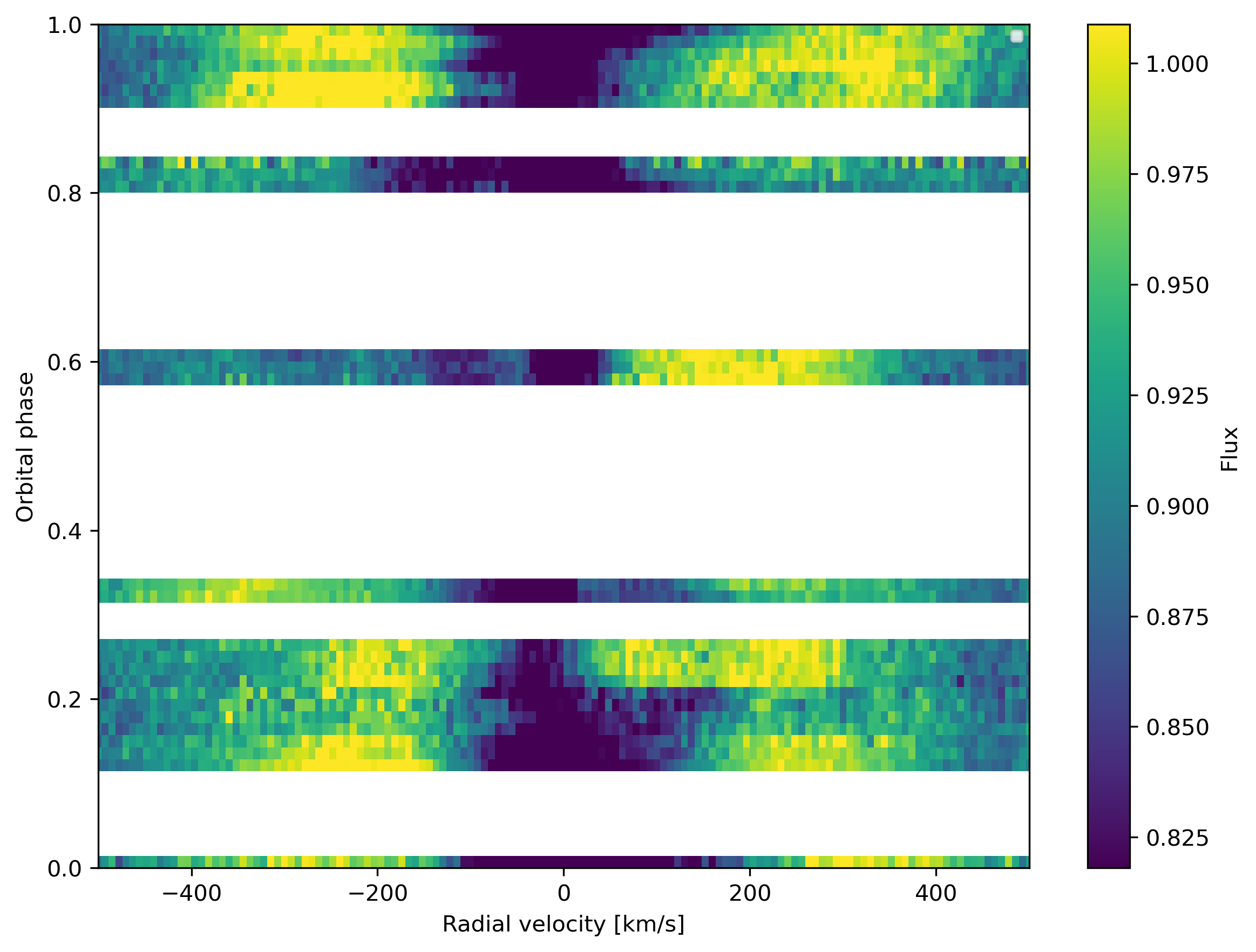}
\caption{
Dynamical spectrum constructed from red-arm spectra centered on the H$\alpha$ region, phase-folded using the 
orbital period derived from the \textit{TESS} light curve. The horizontal axis shows radial velocity (km/s), and the vertical axis represents orbital phase. Dark blue bands correspond to absorption features, while yellow regions indicate relative emission or continuum enhancement.
\label{figure4-}}
\end{figure}

\begin{figure}
\centering
\includegraphics[width=0.5\textwidth]{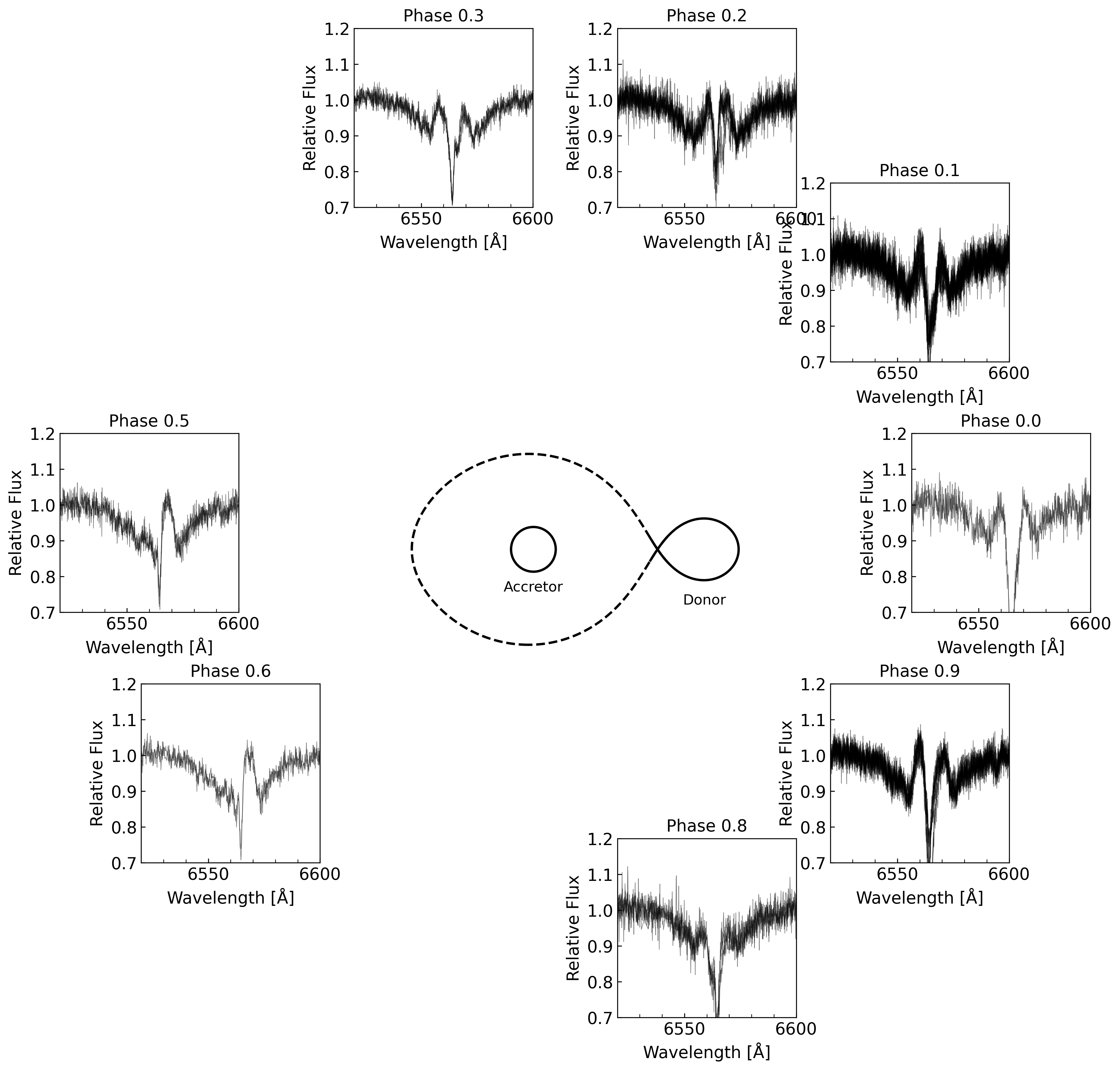}
\caption{
The central panel illustrates the Roche geometry and stellar configurations of the semi-detached binary, while the surrounding panels present the evolution of the H$\alpha$ line profiles across various orbital phases.
\label{figure5}}
\end{figure} 

Double-peaked H$\alpha$ emission has long been recognized as a characteristic of accretion discs in 
interacting Algol-type binaries \citep{Wyse1934, Carpenter1930, AtwoodStone2012, Lehmann2020}. 
\cite{Richards1999} classified such emissions into four morphological types, with double-peaked profiles typically tracing (transient or stable) disc flows, while single-peaked profiles are often linked to gas streams or related accretion phenomena. Systems with $P_\mathrm{orb} < 2.4$\,days generally lack emission, while those with $P_\mathrm{orb} > 6$\,days usually show stable, double-peaked profiles consistent with persistent Keplerian discs \citep{Budaj2005, Zhou2018}. Intermediate systems ($P_\mathrm{orb} \sim 4.5$--6\,days) may show unstable, short-lived discs, 
producing broader, phase-dependent profiles driven by turbulence. Classical discs yield narrower, more stable lines dominated by near-Keplerian velocities \citep{Kaitchuck1989}.

%Double-peaked H$\alpha$ emission has long been recognized as a signature of accretion discs in Algol-type interacting binaries \citep{Wyse1934, Carpenter1930, AtwoodStone2012, Lehmann2020}. \citet{Richards1999} classified such systems into distinct morphological types based on H$\alpha$ profile variations, identifying double-peaked and single-peaked emissions as indicators of disc-like and stream-like structures, respectively. The presence and shape of H$\alpha$ emission also correlate with orbital period: short-period systems ($P_\mathrm{orb} < 2.4$~days) often lack strong emission, while long-period systems ($P_\mathrm{orb} > 6$~days) tend to exhibit stable, double-peaked profiles typical of Keplerian discs \citep{Budaj2005, Zhou2018}. Intermediate-period binaries may show broader, variable features linked to transient disc activity \citep{Kaitchuck1989}.However, at phases $\phi = 0.5$ and $\phi = 0.6$, the emission profile transitions into a single-peaked P~Cygni-type structure, characterized by blue-shifted absorption superimposed on redshifted emission. The presence of a P~Cygni profile suggests the existence of an outflowing component, possibly a wind or jet launched from the vicinity of the accretor or the impact region.

Our binary system, with an orbital period of $P_{\mathrm{orb}} = 2.47757(1)$\,days, is a short-period Algol-type binary that exhibits prominent H$\alpha$ emission. We obtained 60 red-arm spectra centered on the H$\alpha$ region and folded them in the orbital phase. To investigate the velocity structure of the emission, we first constructed a H$\alpha$ dynamical spectrum (Figure~\ref{figure4-}). In this two-dimensional phase–velocity map, absorption appears as dark blue while emission manifests as yellow regions. Notably, asymmetric emission components are observed, with the negative-velocity side peaking near $v \approx -250$ km/s and the positive side extending to about $+300$ km/s. This asymmetry, together with its variation across orbital phase, suggests the presence of extended accretion structures, such as a disc. The orbital phase was divided into 10 equal intervals, but the available spectra only fall within 8 of these bins. The resulting profiles are displayed in a circular layout around a semi-detached binary model (Figure~\ref{figure5}). Most orbital phases show clear double-peaked H$\alpha$ emission, indicative of 
rotating accretion structures. 
%To further investigate the velocity structure of the emission, we constructed a H$\alpha$ dynamical spectrum (Figure~\ref{figure4-}). In this two-dimensional phase-velocity map, absorption appears as dark features while emission manifests as bright regions. Notably, distinct emission components are observed near $v \approx \pm 300 $ km/s, suggesting the presence of extended accretion structures, such as a disc.

To assess the physical origin of the emission, we first ruled out the possibility of interstellar contamination, as the observed velocity shifts vary coherently with the orbital phase. We then consider 
whether the emission could be caused by material ejected from near-critically rotating stars. The critical rotational velocity refers to the maximum speed at which a star can rotate before centrifugal forces at the equator overcome gravitational binding, potentially leading to mass loss. It is defined by the balance between gravitational and centrifugal forces at the stellar equator. Using the measured stellar masses and radii (see Table~\ref{tab:rv}), we compute the critical velocities as
\[
v_{\mathrm{crit,1}} = 523.5 \pm 22.1~\mathrm{km/s}, \quad
v_{\mathrm{crit,2}} = 130.9 \pm 9.0~\mathrm{km/s}.
\]
However, the observed projected rotational velocities ($v \sin i$), which represent the component of the rotational velocity along our line of sight, are significantly lower:
\[
v_{\mathrm{rot,1}} = 20 \pm 9~\mathrm{km/s}, \quad
v_{\mathrm{rot,2}} = 32 \pm 10~\mathrm{km/s} \quad (\text{see Section~\ref{sec:2}}).
\]
This large discrepancy indicates that neither star is rotating fast enough to support the formation of a decretion disc via near-critical rotation.

We then considered the possibility of an accretion disc.  
According to the criterion proposed by \citet{1975AL, HessmanHopp1990}, disc formation requires that the minimum circularization radius—the smallest distance at which the transferred material can stably orbit the accretor—exceeds the radius of the accreting star. For this system, the circularization radius is estimated to be $\sim1.42\,R_{\mathrm{d}}$, where $R_{\mathrm{d}}$ denotes the stellar radius of the accretor. This implies that the mass-transfer stream does not directly strike the star but instead can settle into a stable orbit, allowing a disc to form.  

From theoretical considerations, the disc is expected to extend outward to $\sim2.655\,R_{\mathrm{d}}$. Converting this into physical units, we obtain
\begin{equation}
R_{\mathrm{out,model}} \approx 3.36^{+0.32}_{-0.43}\,R_\odot,
\end{equation}
which corresponds to $\sim65\%$ of the Roche lobe radius ($R_{\mathrm{L1}} = 5.61^{+0.10}_{-0.22}\,R_\odot$).  

Independent observational constraints are available from the H$\alpha$ line profile (Figure~\ref{figure4-}), which displays prominent double-peaked emission. Assuming Keplerian rotation around the primary star, the corresponding outer disc radius is estimated as \citep{Zhou2018}  
\begin{equation}
R_{\mathrm{out,obs}} = \frac{GM_1}{v^2}\approx 3.58 \pm 0.43 \,R_\odot.
\end{equation} 

The close agreement between the theoretical prediction ($R_{\mathrm{out,model}}$) and the spectroscopic estimate 
($R_{\mathrm{out,obs}}$) strongly supports the presence of a stable accretion disc in the system. Both approaches indicate that 
the disc extends from near the stellar surface out to $\sim3.3$–$3.6\,R_\odot$, consistent with expectations for semi-detached binaries. 

\medskip
In the following analysis, we construct a simple accretion disc model following the classical thin-disc framework of \citet{1973Shakura}, using the stellar and binary parameters listed in Table~\ref{tab:physical_params}. The disc is assumed to be geometrically thin, optically thick, and in Keplerian rotation around the primary star.  

The radial extent of the disc is taken to be from an inner boundary at the stellar surface ($R_\mathrm{in} \simeq 1.0\,R_\mathrm{d}$) to an outer boundary $R_\mathrm{out} \simeq 2.655\,R_\mathrm{d}$. The effective temperature distribution is then given by
\begin{equation}
T_{\mathrm{eff}}(R) = \left[ \frac{3GM_1\dot{M}}{8\pi \sigma R^3} 
\left(1 - \sqrt{\frac{R_\mathrm{in}}{R}} \right) \right]^{1/4},
\end{equation}

where $M_1$ is the primary mass, $\dot{M}$ the mass accretion rate, and $\sigma$ the Stefan--Boltzmann constant. The accretion rate was estimated to be $\dot{M} \sim 10^{-10}\,M_\odot\,\mathrm{yr^{-1}}$ based on evolutionary models. Using the observed H$\alpha$ flux and the empirical relation between $L({\rm H}\alpha)$ and $L_{\rm acc}$ \citep{Fang2009,Smallwood2022}, we obtained a mass accretion rate of $\dot{M} \sim 
(2.17-2.26)\times10^{-10}\,M_\odot\,\mathrm{yr^{-1}}$, which is consistent in order of magnitude with the theoretical estimate. The small difference in the coefficient does not significantly affect our theoretical disc modeling; therefore, we retain the value of 
$10^{-10}\,M_\odot\,\mathrm{yr^{-1}}$ in our calculations. At the inner edge, the disc reaches an effective temperature of about $1.05 \times 10^3$\,K.

The total mass of the disc is estimated by integrating the surface density profile $\Sigma(R)$:
\begin{equation}
M_\mathrm{disc} = 2\pi \int_{R_\mathrm{in}}^{R_\mathrm{out}} \Sigma(R)\,R\,dR,
\end{equation}
with
\begin{equation}
\Sigma(R) = \frac{\dot{M}}{3\pi \mu(R)} 
\left[ 1 - \left( \frac{R_0}{R} \right)^{1/2} \right],
\end{equation}
where $R_0$ is the stellar radius, $R$ the disc radius, $\dot{M}$ the accretion rate, 
and $\mu(R)$ the kinematic viscosity coefficient, which varies with $R$ as part of the internal disc structure. We can obtain
\begin{equation}
M_\mathrm{disc} \approx 4.7 \times 10^{-14}\,M_\odot.
\end{equation}
This mass is several orders of magnitude smaller than the stellar mass, demonstrating that the disc’s self-gravity can safely be neglected.

\begin{figure}
\centering
\includegraphics[width=0.5\textwidth]{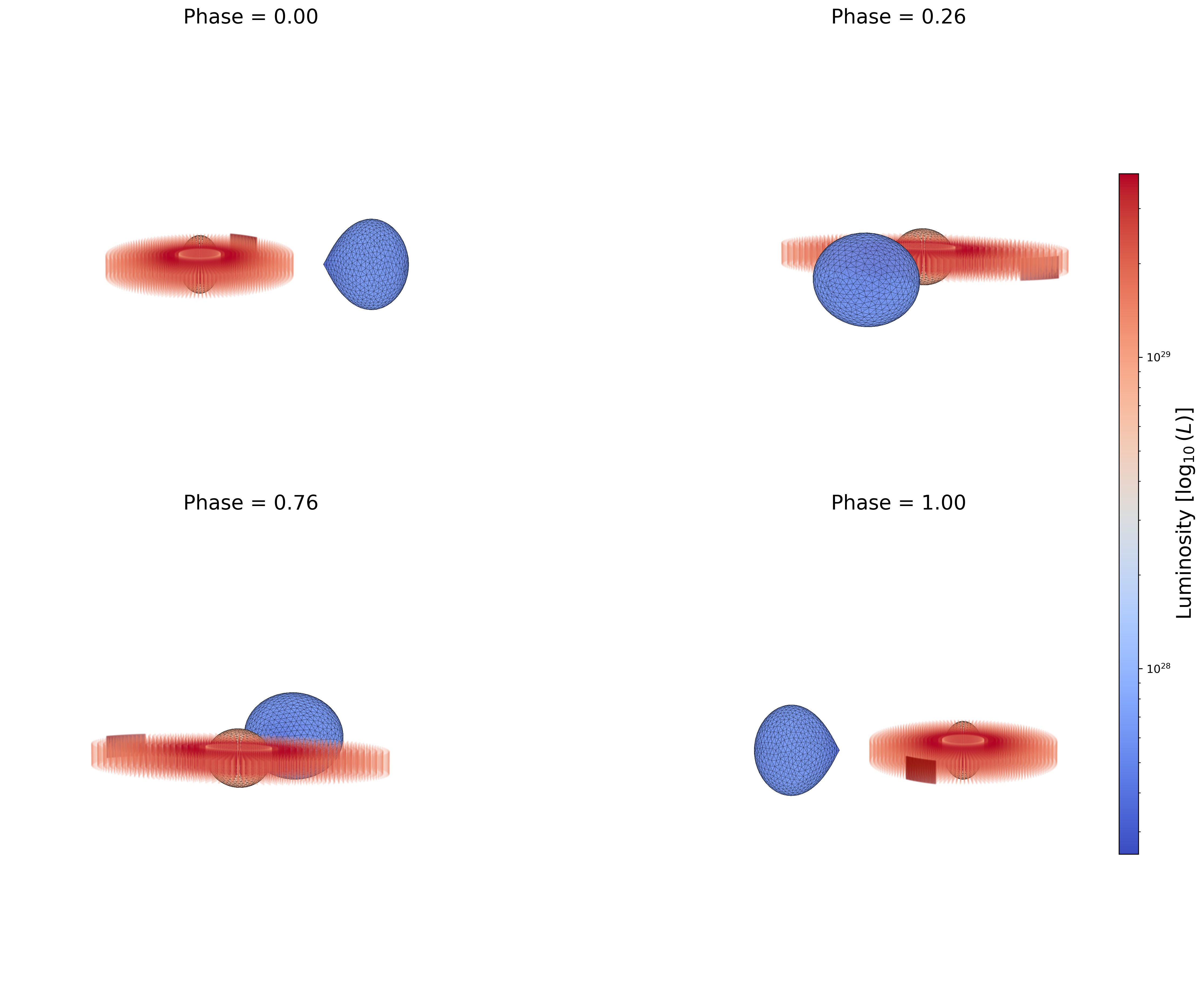}
\caption{Model visualization of the binary system \textit{Gaia} DR3 1892576067672499328 at selected orbital phases. The primary is the accreting star surrounded by an accretion disc, while the secondary acts as the donor star that fills its Roche lobe. The color bar indicates luminosity, with the deep red region marking the location of the hotspot on the disc. For visual clarity, the system is rendered at an inclination angle of $80^\circ$, whereas the actual inclination derived from our modeling is $46.934^\circ$. 
\label{figure8-}}
\end{figure}

Finally, we emphasize the consistency between theory and observation: 
\begin{equation}
R_{\mathrm{out,model}} \approx 3.36^{+0.32}_{-0.43}\,R_\odot,
\qquad
R_{\mathrm{out,obs}} \approx 3.58 \pm 0.43 \,R_\odot.
\end{equation}
The close match between these values provides evidence for the presence of an accretion disc in this system (see Figure~\ref{figure8-}).

%The innermost region of the disc reaches an effective temperature of approximately $2.8\times10^4\,\mathrm{K}$, and the total disc mass is estimated to be $4.55 \times 10^{-10}\,M_\odot$, well below the dynamical mass of the primary, ensuring the disc's self-gravity can be neglected. Tidal forces truncate the disc at approximately $76\%$ of the Roche-lobe radius, in accordance with theoretical expectations. In summary, disc-related parameters are summarized in Table~\ref{tab:velocities}.

%In summary, the low stellar rotation rates exclude the formation of a decretion disc, while the high impact velocity and observed H$\alpha$ line morphology strongly support the presence of a stable, Keplerian accretion disc in this system. For the full set of derived parameters, see Table~\ref{tab:velocities}.

\begin{table}
\centering
\caption{Rotational velocities and accretion disc properties.}
\label{tab:velocities}
\begin{tabular}{lcc}
\hline\hline
Parameter & Value & Unit \\
\hline
\multicolumn{3}{c}{\textit{Rotational velocities}} \\
\hline
Critical rotational velocity (primary) & $523.5 \pm 22.1$ & km/s \\
Critical rotational velocity (secondary) & $130.9 \pm 9.0$ & km/s \\
Observed rotational velocity (primary) & $20 \pm 9$ & km/s \\
Observed rotational velocity (secondary) & $32 \pm 10$ & km/s \\
\hline
\multicolumn{3}{c}{\textit{Emission} features} \\
\hline
%H$\alpha$ emission wing velocity & $\pm300$ & km/s \\
H$\alpha$ mission peak & $-250$ & km/s \\
H$\alpha$ mission peak & $+300$ & km/s \\
\hline
\multicolumn{3}{c}{\textit{Accretion disc structure}} \\
\hline
Inner disc radius $R_{\mathrm{in}}$ & $1.00\,R_\mathrm{d} \approx 1.265$ & $R_\odot$ \\
Outer disc radius $R_{\mathrm{out}}$ & $2.655\,R_\mathrm{d} \approx 3.36$ & $R_\odot$ \\
disc truncation radius & $\sim 65\%$ & $ R_\mathrm{L} $\\
Innest disc temperature & $1.05 \times 10^3$ & K \\
disc mass $M_{\mathrm{disc}}$ & $4.73 \times 10^{-14}$ & $M_\odot$ \\
Density ratio $\rho_{\mathrm{disc}} / \rho_{\mathrm{star}}$ & $2.35 \times 10^{-14}$ & -- \\
Mass ratio $M_{\mathrm{disc}} / M_\mathrm{d}$ & $2.6 \times 10^{-14}$ & -- \\
\hline
\end{tabular}
\end{table}

\subsection{A Possible Third Body}
\label{sec:4.2}

%\begin{figure}[ht!]
%\centering
%\includegraphics[width=11cm]{6.png}
%\caption{Systemic velocity $\gamma$ of the star \textit{Gaia} DR3 1892576067672499328 as a function of Modified Julian Date (MJD). The variation in $\gamma$ suggests the presence of an external perturbation.
%\label{figure 6}}
%\end{figure}

\begin{figure}
\centering
\includegraphics[width=0.5\textwidth]{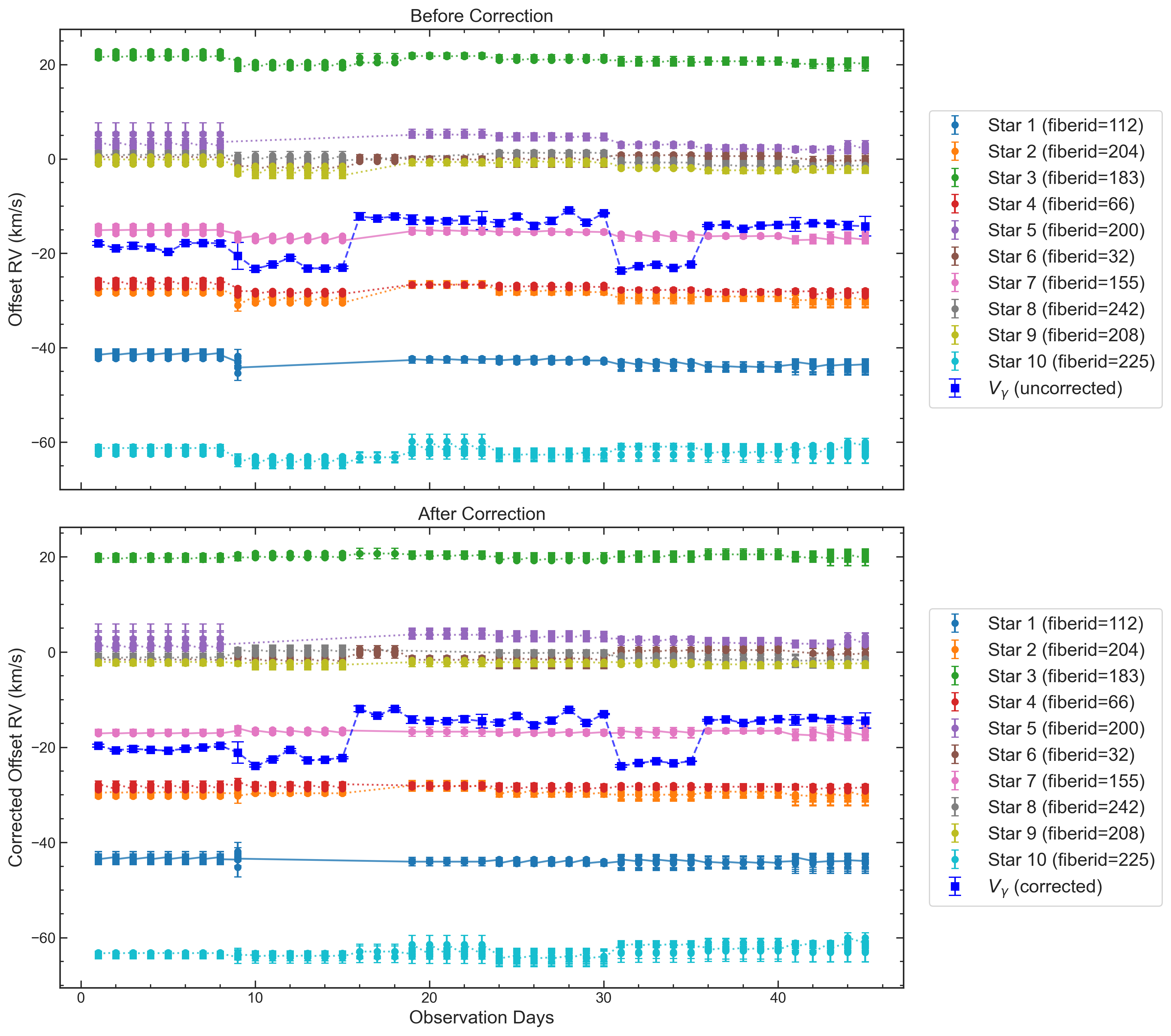}
\caption{
Radial velocity (RV) variations of the 10 stars nearest to the target star \textit{Gaia} DR3 1892576067672499328, observed with LAMOST over multiple epochs. The upper panel shows the raw RV measurements before zero-point correction, while the lower panel displays the results after applying the correction. Each point represents an individual measurement, with stars color-coded and labeled by their fiber ID. Note that the vertical positions of the stars are not their absolute RVs; instead, small vertical shifts have been applied to separate the curves. These offsets are scaled according to the distance of each fiber from the two target fibers (IDs 113 and 152), such that stars located closer to the target are plotted nearer to it, while more distant stars are plotted farther away. Blue squares with dashed lines represent the systemic velocity $V_\gamma$ of the binary model at each epoch, while error bars indicate measurement uncertainties. 
After zero-point correction, the systemic velocity curve becomes significantly more stable.
\label{figure6}}
\end{figure}

During the binary orbit fitting with \texttt{RadVel} (Section~\ref{sec:3.1}), we already applied zero-point 
corrections to the radial velocities using \textit{Gaia} data to ensure consistency across different observing epochs \citep{2022AJWang,2025Patrick}. Despite this correction, the systemic velocity $\gamma$ still shows 
significant variation—exceeding 10 km/s over the observation baseline—which cannot be accounted for by orbital motion in a standard binary system (see Figure~\ref{figure6}). This long-term trend may suggest the presence 
of an additional companion perturbing the binary, potentially forming a hierarchical triple system. Given the 
corrected radial velocities $v_1$ and $v_2$ at different Modified Julian Dates (MJDs), and the mass ratio $q = M_2 / M_1$, we calculated the instantaneous systemic velocity $\gamma$ at each epoch using the relation:

%While fitting the binary orbit with RadVel (Section \ref{sec:3.1}), we found that the systemic velocity $\gamma$ exhibits variations over time that cannot be explained by orbital motion within a two-body framework, suggest the influence of an external star. Specifically, given the radial velocities $v_1$ and $v_2$ at various Modified Julian Dates (MJDs), and the mass ratio $q = M_2 / M_1$, we computed the systemic velocity $\gamma$ for each MJD using the relation:

\begin{equation}
\gamma = \frac{v_1 + q v_2}{1 + q}.
\end{equation}

In a binary system, the systemic velocity $\gamma$ is expected to remain constant. However, as illustrated in Figure~\ref{figure6}, where the horizontal axis represents the number of observations, the measured values for our target star (blue points) exhibit clear jumps, indicating variability in $\gamma$. This behavior is further confirmed in Figure~\ref{figure6-}, which shows the systemic velocity as a function of time; the drift of $\gamma$ with time is evident. Although the current observations do not span a full outer orbital period, the observed trend is broadly consistent with the reflex motion expected from the presence of a tertiary component. To place a lower limit on the properties of such a hypothetical tertiary component, we assume that the outer orbit is circular ($e = 0$) and viewed edge-on ($i = 90^\circ$). We estimate the minimum mass of the third body using the standard mass function:

\begin{equation}
f_{\text{third}} = \frac{M_{3}^3 \sin^3 i_1}{(M_{1} + M_{2}+ M_{3})^2} = \frac{P_{\text{out}} K_{\text{out}}^3}{2\pi G},
\end{equation}

%If the system were an isolated binary, $\gamma$ would remain constant. However, as shown in Figure~\ref{figure 6}, we find a quasi-sinusoidal variation in $\gamma$ over time. This behavior suggests that the center of mass of the binary itself is in motion, likely due to the gravitational influence of a tertiary component in a hierarchical triple configuration. Although our data do not yet span a complete orbit of this outer companion, the observed trend is consistent with reflex motion induced by a distant third body. Assuming the outer orbit is circular ($e_1 = 0$) and viewed edge-on ($i_1 = 90^\circ$), we can estimate the minimum mass of the tertiary component using the mass function:

where $K_{\text{out}}$ is the semi-amplitude of the systemic velocity variation, and $P_{\text{out}}$ is the period of the outer orbit. This edge-on, circular assumption is commonly adopted in the literature as it yields the minimum companion mass consistent with the observed velocity signal \citep{2016Borkovit, 2025AJT}. We note, however, that the inner binary inclination is $i \approx 46^\circ$ from our modeling; if the tertiary orbit is misaligned, Kozai–Lidov oscillations could in principle be induced, potentially affecting both the inner binary and the accretion disc \citep{2016ARA, 2019MNRASF}. Furthermore, if the outer orbit has non-zero eccentricity, the range of possible solutions broadens and the true companion mass would generally be larger than our circular, edge-on estimate. Therefore, the values derived here should be regarded as conservative lower limits. Based on the absence of any clear periodic signal in the observed-minus-calculated 
(O–C) diagram constructed from two \textit{TESS} sectors spanning $\sim25$ days, we conservatively adopt $P_{\rm out} > 50$~days as a lower limit. Using this value and the observed amplitude $K_{\rm out}$, we estimate a minimum mass of $M_3 > 0.369\,M_\odot$ for the potential third body. To assess the dynamical stability of this possible triple system, we applied the widely used criterion based on the ratio of outer to inner binary periods \citep{1995ApJ...455..640E}. In this framework, the critical initial period ratio $X_0^{\rm min}$ is linked to the critical ratio of the outer orbit’s periastron distance to the inner orbit’s apastron distance $Y_0^{\rm min}$ through:

\[
\left(X_{0}^{\rm min}\right)^{2/3} = \left(\frac{1}{1+q_{\rm out}}\right)^{1/3} \frac{1+e_{\rm in}}{1-e_{\rm out}} Y_0^{\rm min}, 
\]

where $Y_0^{\rm min}$ can be approximated by

\[
Y_0^{\rm min} \approx 1 + \frac{3.7}{q_{\rm out}^{-1/3}} + \frac{2.2}{1 + q_{\rm out}^{-1/3}} + \frac{1.4}{q_{\rm in}^{-1/3}} \frac{q_{\rm out}^{-1/3} - 1}{q_{\rm out}^{-1/3} + 1}.
\]

Here, $q_{\rm in}$ and $q_{\rm out}$ are the inner and outer mass ratios, and $e_{\rm in}$ and $e_{\rm out}$ are the eccentricities of the inner and outer binaries, respectively. For the estimated parameters of our system, the outer-to-inner period ratio $P_{\rm out}/P_{\rm in}$ exceeds the critical value, indicating that the system is dynamically stable \citep{Toonen2017, 2019Tokovinin&Moe, 2008eggleton, He2023}.

%where $K_{\text{out}}$ is the amplitude of the systemic velocity variation and $P_{\text{out}}$ is the outer orbital period. Assuming a conservative lower limit of $P_{\text{out}} > 50$ days—given the absence of any periodic signature in the 25-day \textit{TESS} O–C diagram—we estimate a minimum mass of $M_3 \gtrsim 0.378,M_\odot$ for the third body. This estimate ensures dynamical stability, consistent with the commonly adopted stability criterion $P_{\text{out}}/P_{\text{in}} \gtrsim 5$ \citep{1995ApJ...455..640E,Toonen2017,2019Tokovinin&Moe,2008eggleton,He2023}.

%To confirm the presence and nature of the tertiary companion, continued high-precision radial velocity monitoring is essential. Long-term observations, potentially complemented by astrometric or interferometric techniques, will help constrain the outer orbit and further elucidate the architecture of this multiple system.

\subsection{Evolutionary Pathway}\label{sec:4.3}

\begin{figure}
\centering
\includegraphics[width=0.5\textwidth]{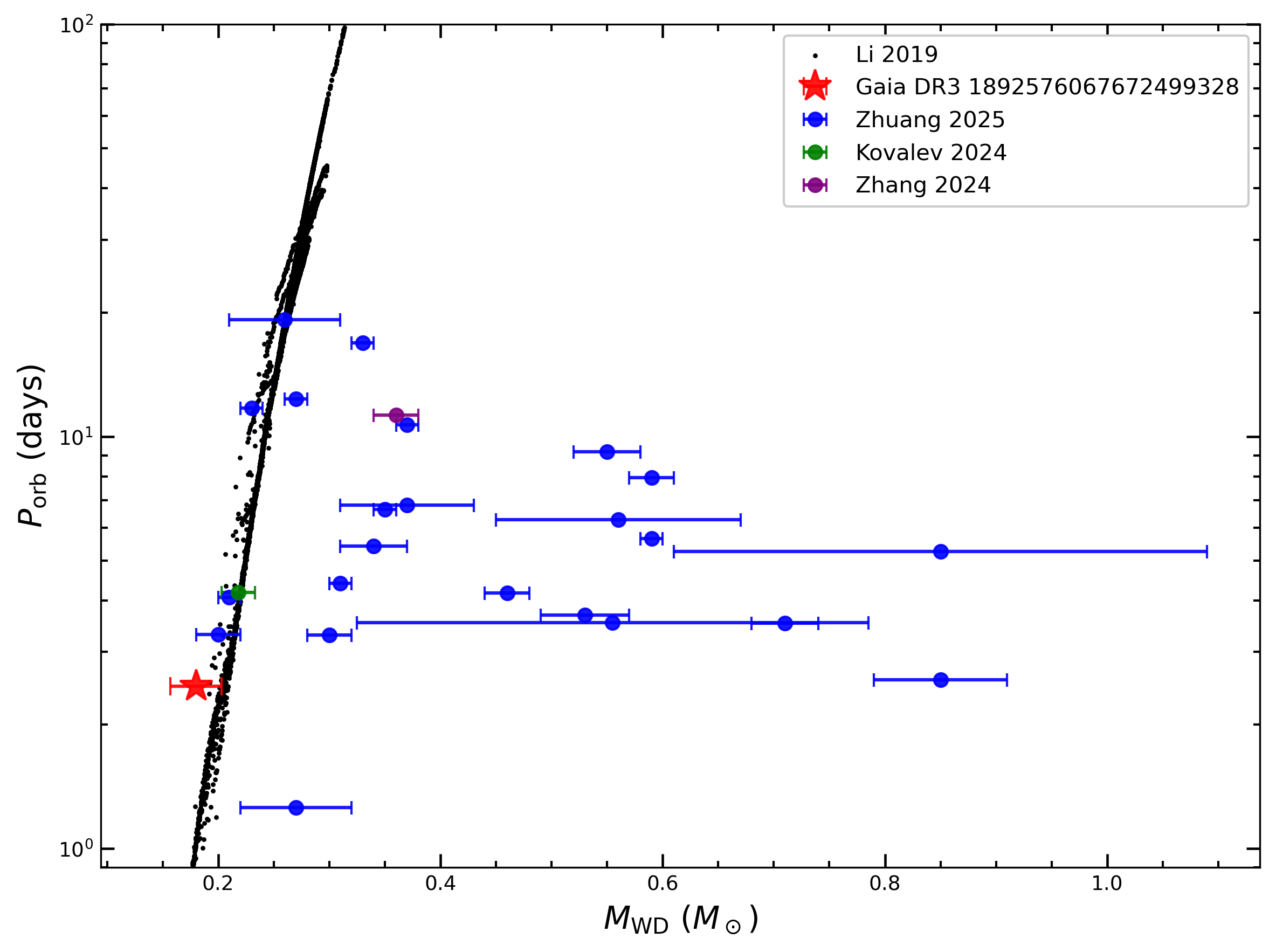}
\caption{Orbital period ($P_\mathrm{orb}$) versus white dwarf mass ($M_\mathrm{WD}$) diagram. The black dots 
represent the predicted distribution from binary evolution models by \citet{2019Li}. The red pentagram marks 
our system, with error bars showing measurement uncertainties. Blue, green, and purple circles correspond to systems reported by \citet{Zhuang2025}, \citet{Kovalev2024}, and \citet{Zhang2024}, respectively. %This plot illustrates the position of our system relative to known white dwarf binaries, suggesting it falls within the typical parameter space of He WD + MS evolutionary channels.
\label{figure7}}
\end{figure}

\begin{figure}
\centering
\includegraphics[width=0.5\textwidth]{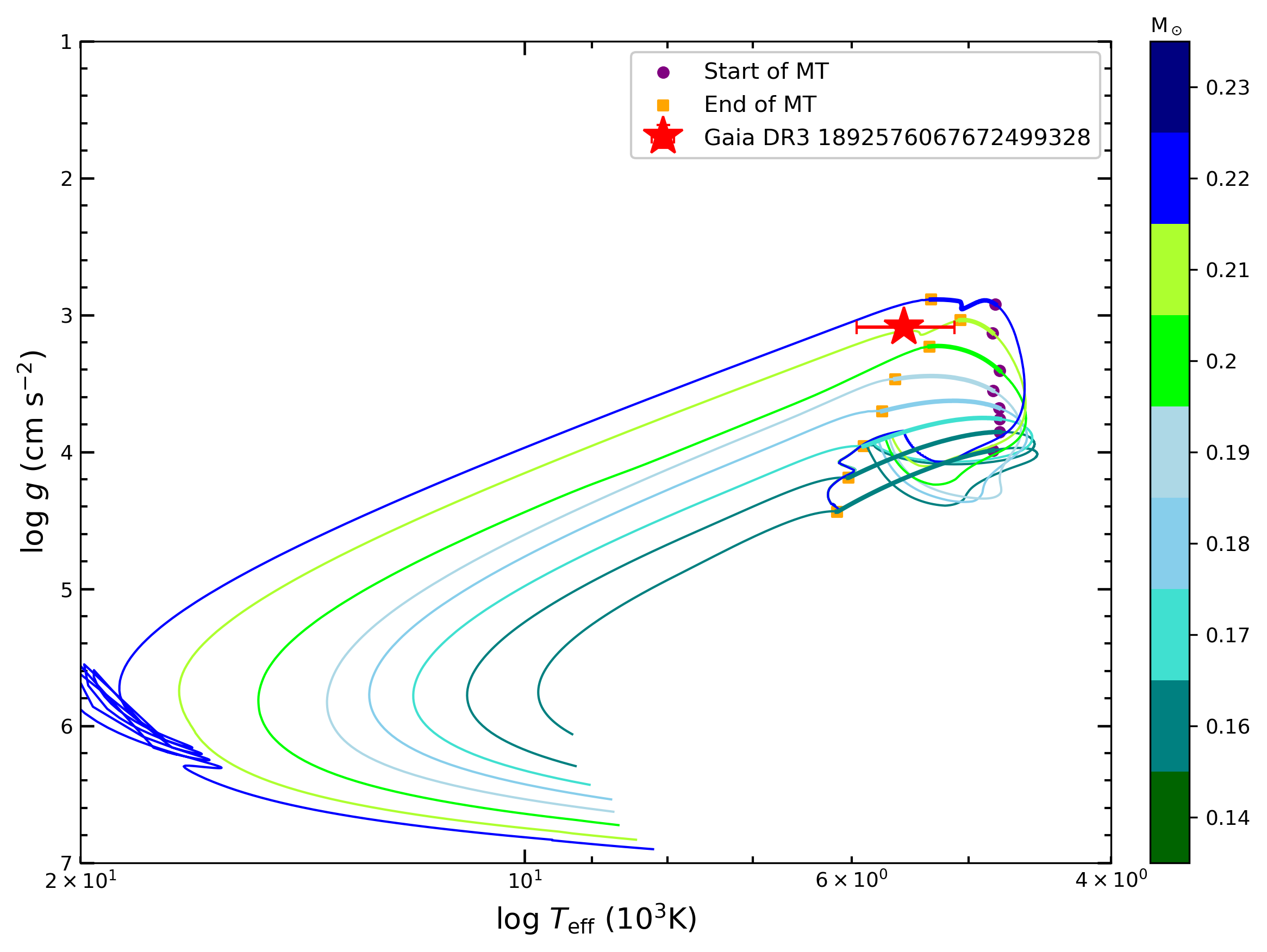}
\caption{Evolutionary tracks of several low-mass stars in the log~$T_{\rm eff}$–log~$g$ plane, based on binary evolution models from \citet{2019Li}. Each track corresponds to a system with different initial parameters, 
and the color denotes the stellar mass, as indicated by the color bar on the right. Bold segments of the 
tracks highlight phases of mass transfer, with blue circles and green squares marking the onset and 
termination of mass transfer, respectively. The red pentagram represents the observed subgiant companion of our binary system.
\label{figure11}}
\end{figure}

The current configuration of Algol-type binary system \textit{Gaia} DR3 1892576067672499328—featuring a 
low-mass, evolved subgiant companion and a relatively short orbital period—suggests a potential evolutionary 
trajectory toward the formation of a helium-core white dwarf (He WD). Such configurations are indicative of 
past mass-transfer episodes, where the initially more massive star has been stripped of most of its envelope, 
leaving behind a low-mass core. This evolutionary scenario is well established in binary stellar evolution theory \citep{ChenHan2003,Istrate2014,2016Istrate,2024Ge}.

To examine this hypothesis, we compare the observed properties of our system with predictions from binary 
evolution models and observational samples of similar short-period binaries. Figure~\ref{figure7} presents the distribution of orbital period versus white dwarf mass ($P_{\rm orb}$–$M_{\rm WD}$). The black dots show the 
model-predicted distribution from \citet{2019Li}, while the red pentagram represents our system. The 
horizontal error bars reflect the uncertainty in the companion’s mass. Also plotted are Algol-type binary 
systems from \citet{Zhuang2025}, \citet{Kovalev2024}, and \citet{Zhang2024}. The position of our system 
overlaps with the region typically populated by EL~CVn-type binaries, which are composed of a low-mass proto-He WD and an A- or F-type main-sequence star \citep{Maxted2014}.

To further assess the evolutionary status of our system, Figure~\ref{figure11} shows the evolutionary tracks 
of several low-mass stars in the $\log T_{\rm eff}$–$\log g$ plane. These tracks are selected from detailed 
binary evolution simulations by \citet{2019Li}, corresponding to stellar masses in the range of $0.14$–$0.23~M_\odot$, which brackets the estimated mass of the donor. This allows us to compare the observed 
position of the donor with theoretical expectations during different stages of mass transfer and nuclear 
evolution. Each track corresponds to a distinct initial mass and orbital configuration, and the color 
indicates the final mass of the stripped star. The bold segments of the tracks mark the phases of mass 
transfer, with blue circles and green squares denoting the onset and cessation of mass transfer, respectively. The red pentagram indicates the observed location of the subgiant companion in our system. Its position is 
consistent with that of stripped-envelope stars nearing the end of their mass-transfer phase, transitioning toward the proto–He WD regime \citep{2016Istrate,2017MNRAS,2021El-Badry}.

Interestingly, the observed location of our system in the $P$–$M_{\rm WD}$ diagram appears to lie slightly 
above several model-predicted evolutionary sequences. This discrepancy underscores certain tensions between 
theoretical expectations and observations. Notably, previous studies have demonstrated that the mass–period 
relation is largely insensitive to the adopted mass-transfer efficiency or angular momentum loss prescriptions \citep{2013Chen}. Instead, this relation is primarily dictated by the donor star’s internal structure at the 
onset of Roche-lobe overflow. As a result, variations in stellar input physics—such as the treatment of 
convection, element diffusion, and opacity—can shift the predicted $P$–$M_{\rm WD}$ relation to a certain 
extent. Additional evidence for this discrepancy comes from the $T_{\rm eff}$–$\log g$ diagram 
(Figure~\ref{figure11}). The observed position of the subgiant companion aligns best with evolutionary tracks 
corresponding to a stripped donor star with a final mass of approximately 0.21~$M_\odot$, which is slightly 
higher than the dynamically measured value. This difference is likely attributable to uncertainties in the modeling of stellar structure and evolution. % For instance, the inclusion of element diffusion tends to lower the final donor mass while increasing the orbital separation, potentially reconciling this small mismatch.

Taken together, these comparisons support the interpretation that our binary system is on a typical 
evolutionary pathway toward forming a helium-core white dwarf. Although the ultimate outcome depends on the 
system's ongoing orbital and angular momentum evolution, the current configuration aligns with theoretical 
expectations for the precursors of EL~CVn systems and may provide further observational constraints on their 
formation channel. And such system may offer valuable insights into the input parameters governing stellar structure and binary evolution.

\subsection{Circumbinary disc as a possible alternative explanation}\label{sec:4.4}

In addition to the tertiary companion hypothesis, the observed long-term velocity variations might also be influenced by the presence of a circumbinary disc. Theoretical studies suggest that, in Algol-type binaries, part of the mass transferred from the donor star may not be fully accreted by the gainer, but instead could accumulate in a circumbinary structure \citep{1993MNRAS,2006ApJC}. The gravitational potential of such a disc may perturb the inner binary’s orbit and thus give rise to changes in the systemic velocity or orbital period. Observational evidence from several Algol-type systems also indicates complex orbital period variations, which have been interpreted in terms of long-term angular momentum loss and possibly the presence of circumbinary material \citep{2000AJQ,2000AQ,2001AJQ}.

For this star system, it cannot be ruled out that a circumbinary disc contributes to the observed velocity changes. However, the character of the variations seems somewhat more naturally accounted for by the tertiary companion hypothesis, whereas perturbations caused by a circumbinary disc are generally expected to be slower and less regular. Continued long-term monitoring, particularly combining radial velocity observations with searches for possible infrared excess, would be valuable to further test the circumbinary disc hypothesis.

\section{Summary}\label{sec:5}
We present a comprehensive study of the short-period Algol-type binary \textit{Gaia} DR3 1892576067672499328, 
combining photometric and spectroscopic data. Medium-resolution LAMOST spectra clearly show SB2 features, and 
together with \textit{TESS} light curves, we determine an accurate orbital period of $P = 2.47757 (1)$ days. 
The system comprises a low-mass, Roche-lobe filling secondary ($M_2 = 0.179 ^{+0.011}_{-0.020} \,M_\odot$) 
orbiting an A-type primary ($M_1 = 1.817^{+0.106}_{-0.202} \,M_\odot$), with a mass ratio of $q = 0.098 \pm 0.002$. The asymmetric double-hump shape in the \textit{TESS} light curve suggests surface inhomogeneities. 
Modeling with PHOEBE~2.3 including a hotspot on the primary star reproduces the observed features and yields well-constrained system parameters. We got the parameters to be: $a = 9.697^{+0.172}_{-0.385}\ R_\odot$, 
$R_1 = 1.265^{+0.121}_{-0.160}\ R_\odot$, 
$T_2/T_1 = 0.712^{+0.048}_{-0.031}$, 
$L_3 = 0.205^{+0.057}_{-0.051}$, 
$i = 46.934^{+2.613}_{-1.110}\ ^\circ$, 
$q = 0.098 \pm 0.002$, 
$M_1 = 1.817^{+0.106}_{-0.202}\ M_\odot$, 
hotspot temperature ratio $T_\mathrm{spot}/T_1 = 1.43^{+0.050}_{-0.040}$, 
hotspot radius $= 11.393^{+1.035}_{-0.616}\ ^\circ$, 
and hotspot longitude $= 127.504^{+1.480}_{-0.529}\ ^\circ$.

Red-arm spectra reveal double-peaked H$\alpha$ emission, consistent with the presence of a Keplerian accretion disc structure around the primary star. The shape and velocity shift of the emission lines supports ongoing 
mass transfer. The standard accretion disc model we constructed well conforms to the observational data and 
presents a simulated accretion disc structure of this star. We also detect a long-term variation in the 
systemic velocity $\gamma$, indicating a tertiary component. Assuming a coplanar outer orbit with $P_{\text{out}} > 50$ days, the minimum mass of the third body is $M_3 > 0.369\,M_\odot$, suggesting the system is a  hierarchical triple.

Finally, the current location of the system in the $P_{\rm orb}$–$M_{\rm WD}$ plane (Figure~\ref{figure7}) 
overlaps with predictions from binary evolution models that produce helium-core white dwarfs via stable mass 
transfer. Its companion's position in the $\log T_{\rm eff}$–$\log g$ diagram (Figure~\ref{figure11}) is 
consistent with a stripped subgiant transitioning into a proto-He white dwarf. These results suggest that the 
system is on an evolutionary path toward becoming an EL~CVn-type binary, with its future evolution shaped by 
continued mass transfer and angular momentum loss. This system provides a valuable case study for 
understanding the formation of EL~CVn-type binaries and the late-stage evolution of low-mass Algol-type systems.

\section*{Acknowledgements}

We would like to express our gratitude to Mingkuan Yang, Jian Mou, Lifu Zhang, Chengyuan Wu, Ziqi Zhao and Yuchen Bao for their help. This work is supported by the National Natural Science Foundation of China (NSFC, Nos. 12288102, 12090040/3, 
12125303, 12173081, 12373037, 12403040, 12303106,12525304, 12473034, 12473033, 12333008), the National Key R\&D Program of China (No. 2021YFA1600403/1), the Yunnan Fundamental Research Projects (Nos. 202401BC070007, 
202401AT070139, 202201AT070180, 202501CF070016, 202201BC070003), the Yunnan Revitalization Talent Support Program: Science and Technology Champion Project (No. 202305AB350003), the International Centre of Supernovae, Yunnan Key 
Laboratory (No. 202302AN360001), the Postdoctoral Fellowship Program of CPSF (No. GZC20232976), the Yunnan Youth Talent Project, and the Yunnan Ten Thousand Talents Plan Young \& Elite Talents Project. Jiao Li is supported by the Young Talent Project of Yunnan Revitalization Talent Support Program.

Guoshoujing Telescope (LAMOST) is a National Major Scientific Project built by the Chinese Academy of Sciences and funded by the National Development and Reform Commission. It is operated and managed by the National Astronomical Observatories, Chinese Academy of Sciences.

This paper includes data collected by the \textit{TESS} mission, funded by NASA’s Explorer Program, and also makes use of data from the European Space Agency (ESA) mission \textit{Gaia} (\url{https://www.cosmos.esa.int/gaia}; \url{https://gea.esac.esa.int/archive/documentation/GDR3/index.html}). We also acknowledge the use of the following software: LASP pipeline \citep{2011wu,2015luo}, ULySS \citep{2009kol}, SPECTRUM \citep{1994gray}, and laspec \citep{2020zhangbo,2021zhangbo}.

\section*{DATA AVAILABILITY}
The data underlying this article will be shared on reasonable request to the corresponding author.
%\end{acknowledgments}
%This work is supported by the National Natural Science Foundation of China under Grant Nos.12288102, 12125303, 12090040/3, 12373037, 12403040, 12303106, the National Key R\&D Program of China (Nos. 2021YFA1600403/1 and 2021YFA1600400), the Natural Science Foundation of Yunnan Province (Nos. 202201BC070003, 202001AW070007), the International Centre of Supernovae, Yunnan Key Laboratory (No. 202302AN360001), the "Yunnan Revitalization Talent Support Program"-Science and Technology Champion Project (No. 202305AB350003) and the Postdoctoral Fellowship Program of CPSF (No.GZC20232976). Guoshoujing Telescope (LAMOST) is a National Major Scientific Project built by the Chinese Academy of Sciences. Funding for the project has been provided by the National Development and Reform Commission. LAMOST is operated and managed by the National Astronomical Observatories, Chinese Academy of Sciences. This paper includes data collected by the $TESS$ mission. The TESS mission is funded by NASA’s Explorer Program. This work has also made use of data from the European Space Agency (ESA) mission $Gaia$ (\url{https://www.cosmos.esa.int/gaia}, \url{https://gea.esac.esa.int/archive/documentation/GDR3/index.html}).

%\appendix

%\section{Appendix information}

%\section{Gold Open Access}

%\section{Author publication %charges} \label{sec:pubcharge}
%\input{hty-.bbl}
\bibliography{reference}{}
\bibliographystyle{mnras}

\appendix
\section{Radial Velocity Measurements}

% 重定义表格编号格式为 “A1”, “A2” ...
\renewcommand{\thetable}{\thesection\arabic{table}}
\setcounter{table}{0}

\begin{table*}
\caption{Radial velocity measurements of the star \textit{Gaia} DR3 1892576067672499328. Columns show observation date (MJD), SNR ratio, radial velocities ($v_1$, $v_2$), cross-correlation amplitudes ($a_1$, $a_2$), and their uncertainties. Units for velocities are in km/s.}
\label{tab:rv}
\centering
\begin{tabular}{rrrrrrrrrr}
\toprule
\textbf{MJD} & \textbf{SNR} & \textbf{$v_1$} & \textbf{$v_2$} & \textbf{$a_1$} & \textbf{$a_2$} & \textbf{$\sigma_{v_1}$} & \textbf{$\sigma_{v_2}$} & \textbf{$\sigma_{a_1}$} & \textbf{$\sigma_{a_2}$} \\
\midrule

58421.45417 & 24.31 & -28.30 & 88.40 & 0.418 & 0.262 & 3.34 & 3.30 & 0.0030 & 0.00050 \\
58421.47083 & 24.01 & -29.50 & 89.00 & 0.423 & 0.257 & 3.19 & 3.47 & 0.0010 & 0.00050 \\
58421.49167 & 21.21 & -29.40 & 94.20 & 0.402 & 0.223 & 3.60 & 3.48 & 0.00050 & 0.00050 \\
58421.51042 & 24.04 & -30.20 & 98.50 & 0.406 & 0.269 & 3.075 & 4.13 & 0.0010 & 0.0030 \\
58421.52708 & 28.80 & -31.00 & 95.80 & 0.484 & 0.285 & 3.075 & 3.29 & 0.0010 & 0.0030 \\
58421.54306 & 28.34 & -29.60 & 102.30 & 0.483 & 0.267 & 3.24 & 3.69 & 0.0010 & 0.0030 \\
58421.55972 & 30.51 & -29.70 & 103.70 & 0.515 & 0.289 & 3.12 & 3.30 & 0.0010 & 0.0010 \\
58421.57292 & 21.53 & -29.70 & 103.10 & 0.426 & 0.230 & 3.41 & 3.029 & 0.0010 & 0.0030 \\
58792.45972 & 44.38 & -11.40 & -83.30 & 0.650 & 0.299 & 5.40 & 4.30 & 0.074 & 0.016 \\
58795.45139 & 48.03 & -32.20 & 67.70 & 0.650 & 0.306 & 3.28 & 3.67 & 0.00050 & 0.0010 \\
58795.46736 & 40.17 & -31.40 & 70.00 & 0.609 & 0.288 & 3.19 & 3.44 & 0.00050 & 0.0030 \\
58795.48403 & 39.72 & -30.30 & 74.70 & 0.600 & 0.300 & 3.25 & 3.43 & 0.0050 & 0.0010 \\
58795.50000 & 34.42 & -32.80 & 75.10 & 0.540 & 0.281 & 3.082 & 3.21 & 0.00050 & 0.0030 \\
58795.51667 & 40.95 & -33.30 & 80.20 & 0.604 & 0.309 & 3.18 & 3.37 & 0.00050 & 0.00050 \\
58795.53264 & 33.69 & -33.50 & 84.40 & 0.567 & 0.267 & 3.23 & 3.40 & 0.00050 & 0.00050 \\
59127.55833 & 36.32 & -22.90 & 97.20 & 0.563 & 0.250 & 3.67 & 4.90 & 0.0040 & 0.00050 \\
59127.57431 & 45.54 & -22.90 & 92.80 & 0.619 & 0.268 & 3.19 & 3.31 & 0.0040 & 0.0010 \\
59127.59514 & 42.93 & -23.00 & 98.30 & 0.593 & 0.273 & 3.27 & 3.30 & 0.00050 & 0.0015 \\
59146.50694 & 45.69 & -1.00 & -134.00 & 0.647 & 0.274 & 3.88 & 3.51 & 0.058 & 0.041 \\
59146.52361 & 39.15 & -1.20 & -134.00 & 0.608 & 0.253 & 3.052 & 3.62 & 0.046 & 0.050 \\
59146.53958 & 39.40 & -1.70 & -129.80 & 0.618 & 0.243 & 3.35 & 3.78 & 0.0070 & 0.016 \\
59146.55556 & 30.58 & -1.70 & -128.00 & 0.548 & 0.225 & 3.71 & 4.60 & 0.061 & 0.041 \\
59146.57222 & 20.53 & -1.60 & -129.90 & 0.460 & 0.172 & 4.60 & 4.30 & 0.041 & 0.026 \\
59157.44931 & 48.02 & -25.90 & 111.80 & 0.621 & 0.263 & 3.61 & 3.85 & 0.00050 & 0.00050 \\
59157.46597 & 47.03 & -24.40 & 112.60 & 0.607 & 0.267 & 3.41 & 3.037 & 0.0015 & 0.0015 \\
59157.48194 & 37.96 & -26.50 & 111.90 & 0.561 & 0.242 & 3.55 & 3.029 & 0.0020 & 0.0020 \\
59157.49792 & 44.22 & -25.40 & 112.30 & 0.593 & 0.279 & 3.62 & 3.071 & 0.00050 & 0.0020 \\
59157.51458 & 38.66 & -23.50 & 118.00 & 0.561 & 0.239 & 3.16 & 3.76 & 0.00050 & 0.0010 \\
59157.53056 & 35.30 & -26.10 & 115.50 & 0.540 & 0.241 & 3.37 & 3.074 & 0.0060 & 0.0010 \\
59157.54653 & 33.08 & -23.60 & 111.90 & 0.513 & 0.235 & 3.097 & 3.46 & 0.0010 & 0.00050 \\
59509.49167 & 50.40 & -35.00 & 92.60 & 0.706 & 0.270 & 3.29 & 3.017 & 0.0050 & 0.00050 \\
59509.50694 & 51.92 & -34.00 & 92.50 & 0.715 & 0.281 & 3.13 & 3.035 & 0.00050 & 0.0030 \\
59509.52222 & 46.14 & -33.60 & 92.10 & 0.678 & 0.269 & 3.14 & 3.39 & 0.0010 & 0.00050 \\
59509.53750 & 46.46 & -33.50 & 83.40 & 0.698 & 0.269 & 3.031 & 3.058 & 0.00050 & 0.0030 \\
59509.55278 & 46.05 & -32.80 & 84.40 & 0.692 & 0.266 & 3.0085 & 3.16 & 0.00050 & 0.00050 \\
59511.48472 & 35.83 & -24.20 & 88.70 & 0.635 & 0.253 & 3.16 & 3.66 & 0.00050 & 0.00050 \\
59511.50000 & 34.74 & -24.20 & 91.00 & 0.635 & 0.254 & 3.017 & 3.059 & 0.00050 & 0.00050 \\
59511.51528 & 30.72 & -25.00 & 89.70 & 0.590 & 0.237 & 3.012 & 3.92 & 0.0010 & 0.0030 \\
59511.53056 & 30.34 & -24.80 & 94.90 & 0.597 & 0.230 & 3.069 & 3.65 & 0.0010 & 0.0030 \\
59511.54583 & 27.83 & -24.60 & 94.70 & 0.561 & 0.197 & 3.072 & 3.21 & 0.0020 & 0.00050 \\
59532.44931 & 41.29 & -5.20 & -102.70 & 0.620 & 0.206 & 4.20 & 4.00 & 0.012 & 0.052 \\
59532.46458 & 38.55 & -4.50 & -106.20 & 0.646 & 0.217 & 3.10 & 3.94 & 0.053 & 0.074 \\
59532.49167 & 45.63 & -3.90 & -113.60 & 0.715 & 0.221 & 3.090 & 4.2 & 0.086 & 0.060 \\
59532.50694 & 47.26 & -3.70 & -120.50 & 0.714 & 0.233 & 3.76 & 3.40 & 0.093 & 0.011 \\
59532.52222 & 47.91 & -3.80 & -121.00 & 0.701 & 0.262 & 4.80 & 4.20 & 0.071 & 0.018 \\

%Through the CCF analysis of the system, we obtained several parameters listed in table \ref{tab1}, which includes the RV (km/s) of each component in the system, and the amplitude $A$ after CCF processing.

\bottomrule
\end{tabular}
\end{table*}

\renewcommand{\thefigure}{A\arabic{figure}}
\setcounter{figure}{0}

\begin{figure}
\centering
\includegraphics[width=0.5\textwidth]{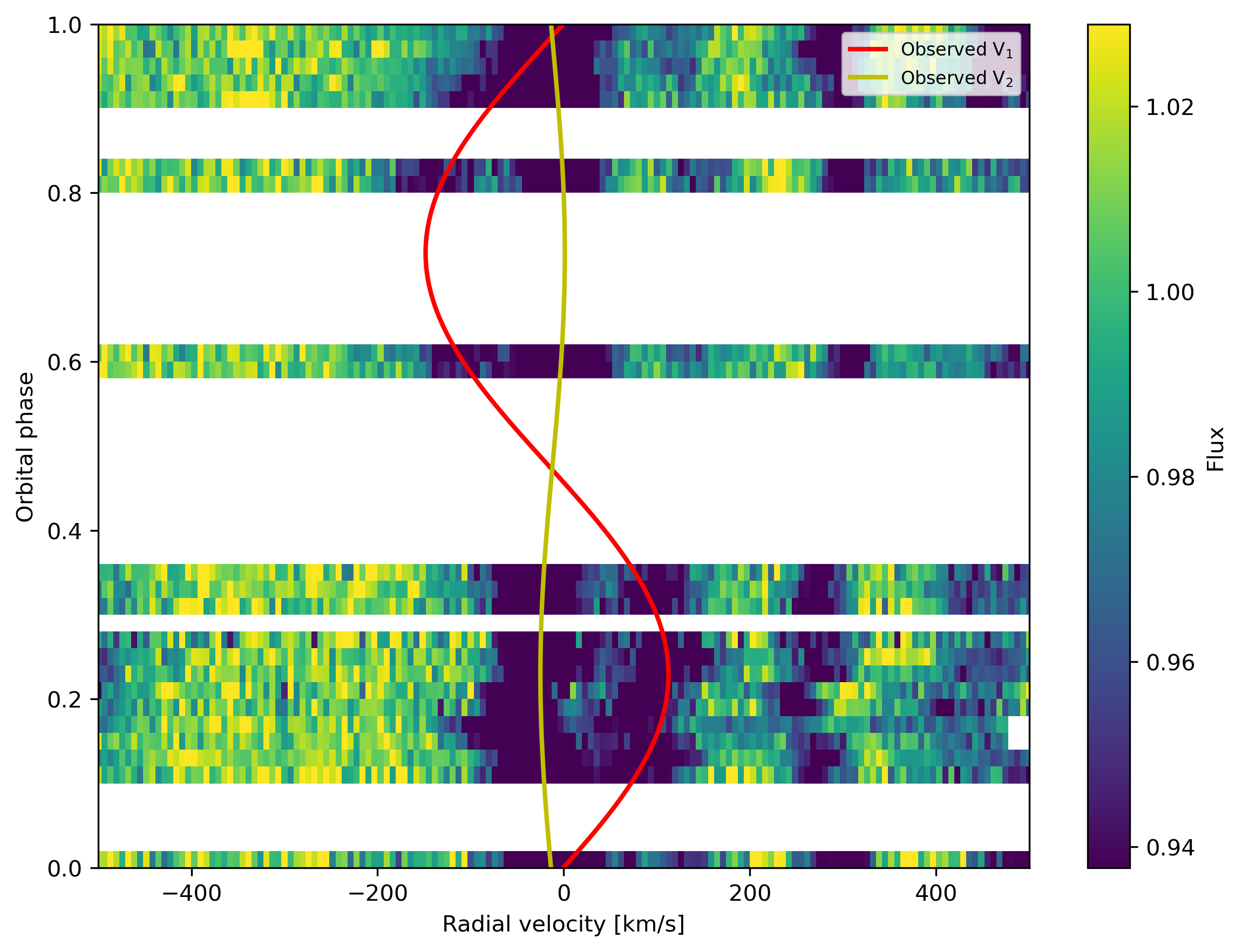}
\caption{The dynamical spectrum shown here is constructed from the blue-arm spectral region around the Mg\text{b} triplet lines. Dynamical spectrum constructed by phase-folding the time-series spectra using the orbital period from the \textit{TESS} light curve. The horizontal axis shows radial velocity (km/s), and the vertical axis indicates orbital phase (bottom to top). Dark bands trace absorption features. Two anti-symmetric absorption tracks, varying sinusoidally with phase. The red and yellow curves show the best-fit RV solutions for the primary and secondary star, respectively.
\label{figure2-}}
\end{figure}

\begin{figure}
\centering
\includegraphics[width=0.5\textwidth]{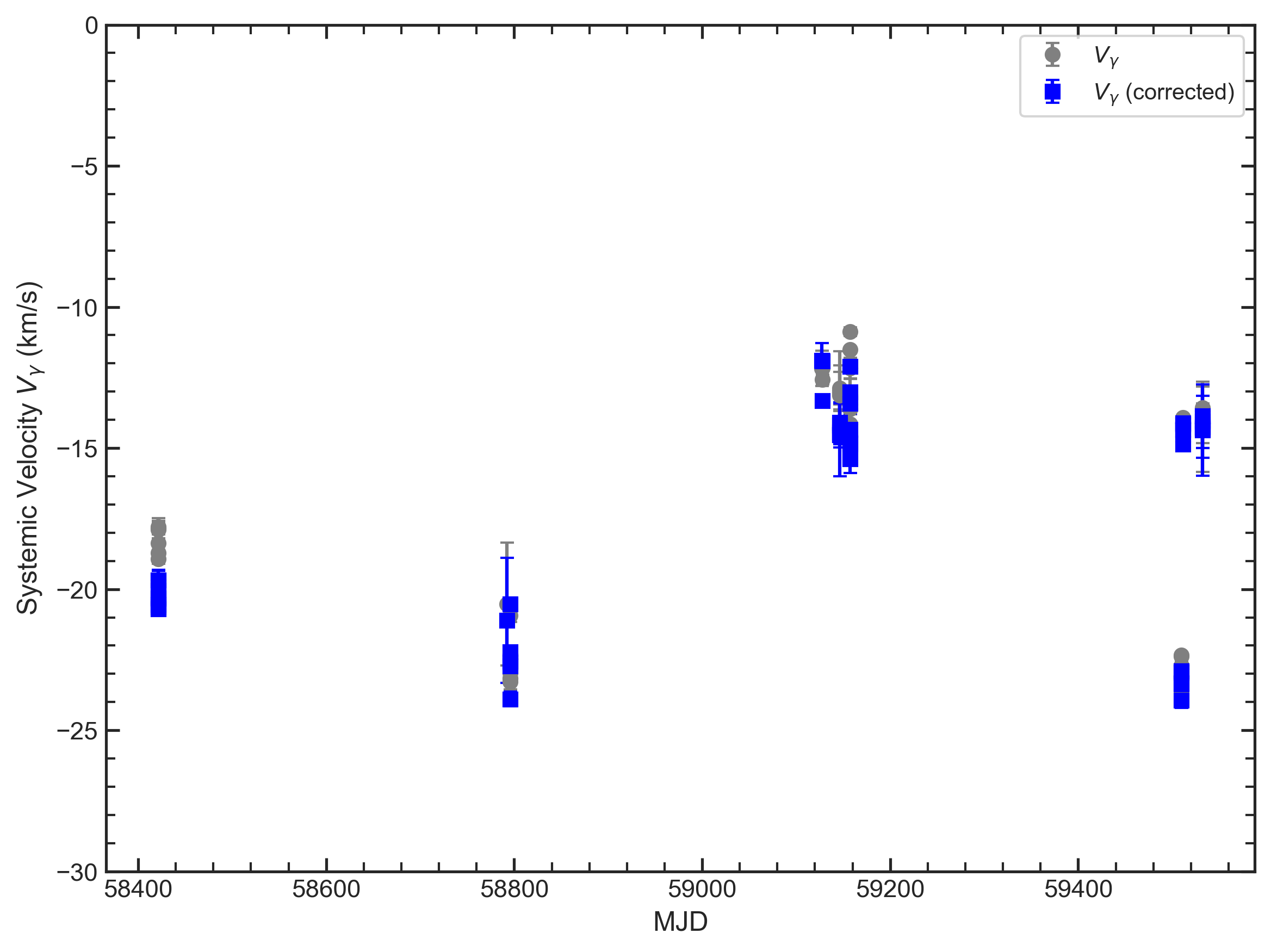}
\caption{
Variations in the systemic velocity ($V_\gamma$) of the system (\textit{Gaia} DR3 1892576067672499328) over time. Gray circles indicate the raw $V_\gamma$ values derived from the radial velocity measurements, while blue squares represent the corrected values after applying the zero-point adjustment based on Gaia DR3 systemic velocities. Error bars denote the measurement uncertainties. A noticeable shift in $V_\gamma$ suggests a possible long-term trend or the presence of an additional body in the system.
\label{figure6-}}
\end{figure}

\label{lastpage}

\end{document}